\documentclass[preprint]{elsarticle}
\usepackage{graphicx,amsmath,amsfonts,algorithm}
\usepackage{graphics}
\usepackage{amssymb}
\usepackage{color}
\usepackage{bbm}
\usepackage[all]{xy}\CompileMatrices

\newcommand{\tr}{{\textrm{tr}}}
\newcommand{\<}{\langle}

\renewcommand{\>}{\rangle}
 
\newcommand{\cplane}[2]{(#1,#2)}
\def\qed{$\Box$}
\def\sem#1{[\hspace{-.35ex}[#1]\hspace{-.35ex}]}
\def\ens#1{\{#1\}}
\def\ie{\textit{i.e.}}
\def\eg{\textit{e.g.}}

\def\dag{^\dagger}
\def\tr#1{\textcolor{red}{#1}}

\def\al{\alpha}
\def\ba{\beta} 

\def\rar{\rightarrow}

\def\Rar{\Rightarrow}

\def\lrar{\longrightarrow}

\def\slar#1{\stackrel{#1}{\lrar}}

\def\mcl{\mathcal} 
\def\mbb{\mathbb} 
 
\def\mfr{\mathfrak}
\newtheorem{prop}{Proposition}\def\PRO{\begin{prop}}\def\ORP{\end{prop}}
\newtheorem{coro}{Corollary}\def\COR{\begin{coro}}\def\ROC{\end{coro}}
\newtheorem{theorem}{Theorem}\def\TH{\begin{theo}}\def\HT{\end{theo}}
\def\TH{\begin{theo}}\def\HT{\end{theo}}
\newtheorem{definition}[prop]{Definition}\def\DE{\begin{defi}}\def\ED{\end{defi}}
\newtheorem{lemma}[prop]{Lemma}\def\LE{\begin{lemme}}\def\EL{\end{lemme}}
\newcommand{\AR}[2][c]{$$\begin{array}[#1]{lllllllllllllll}#2\end{array}$$}
\def\MA#1{\left(\begin{matrix}#1\end{matrix}\right)}
\def\EQ#1{\begin{eqnarray}#1\end{eqnarray}}
\def\hil#1{\mfr H_{#1}}

\def\bra#1{\langle#1{|}}
\def\ctR{\mathop{\wedge}\hskip-.4ex} 
\def\ctwo{{\mbb C}^2}  
\def\ztwo{{\mbb Z}_2}
\def\ost{\frac1{\sqrt2}}

\def\Cx#1{\cx{#1}{}}
\def\Cz#1{\cz{#1}{}}
\def\Cz#1{\cs{#1}{}}
\def\cs#1#2{S_{#1}^{#2}}
\def\cz#1#2{Z_{#1}^{#2}}
\def\cx#1#2{X_{#1}^{#2}}
\def\ei#1{e^{i#1}}

\def\cx#1#2{X_{#1}^{#2}}

\def\mLR#1#2#3#4{{}_{#4}[{M}_{#2}^{#1}]^{#3}}
\def\mR#1#2#3{\mLR{#1}{#2}{#3}{}}

\def\m#1#2{{M}_{#2}^{#1}}
\def\M#1#2{{M}_{#2}^{#1}}
\def\MS#1#2#3#4{{}^{#4}[{M}_{#2}^{#1}]^{#3}}
\def\et#1#2{E_{#1#2}}
\def\etil#1#2{\widetilde{E}_{#1#2}}

\def\CO#1{A_{#1}}
\def\ss#1#2{F_{#1}^{#2}}

\def\GM#1#2#3{{M}_{#3}^{#1,#2}}
\def\GMLR#1#2#3#4#5{{}_{#5}[{M}_{#3}^{#1,#2}]^{#4}}

\def\oqb#1{\ket{\hskip-.1ex+_{#1}}}

\def\oqbn#1{\ket{\hskip-.1ex-_{#1}}}
\def\oqbb#1{\bra{\hskip-.1ex+_{#1}}}
\def\oqbnb#1{\bra{\hskip-.1ex-_{#1}}}

\def\<{\langle}
\def\>{\rangle}

\def\KO{K} 
\def\tr{\mbox{tr}}

\def\ctRZSWAP{\text{SWAP} \ctR Z }

\def\al{\alpha}
\def\ba{\beta} 
\def\ga{\gamma}
\let\da\delta

\def\ta{\theta}

\def\Ga{\Gamma}

\def\vec#1{{\bf #1}}

\def\ket#1{| #1 \rangle}

\def\bra#1{\langle #1 |}

\def\proj#1{\ket{#1}\bra{#1}}

\def\norm#1{\mid\mid #1 \mid\mid}

\def\tr{\text{Tr}}

\newcommand{\inp}[2]{\langle{#1}\mid {#2}\rangle}

\newcommand{\hilb}{\mathfrak H}
\newcommand{\adj}[1]{#1^{\dagger}}

\definecolor{Ocolor}{rgb}{.9,.1,.1}
\definecolor{Bcolor}{rgb}{.1,.1,.9}

\begin{document}

\title{Ancilla-driven quantum computation with\\ twisted graph states} 

\author[inf]{E. Kashefi}
\author[str]{D. K. L. Oi}
\author[ucl]{D. E. Browne}
\author[ucl]{J. Anders}
\author[hw]{E. Andersson}

\address[ucl]{Department of Physics \& Astronomy, University College London, London, UK}
\address[hw]{School of Engineering and Physical Sciences, Heriot-Watt University, Edinburgh, UK}
\address[inf]{School of Informatics, University of Edinburgh, Edinburgh, UK}
\address[str]{Department of Physics, University of Strathclyde, Glasgow, UK}

\begin{abstract} 
We introduce a new paradigm for quantum computing called Ancilla-Driven Quantum Computation (ADQC) combines aspects of the quantum circuit~\cite{Deutsch89} and the one-way model~\cite{RB01} to overcome challenging issues in building large-scale quantum computers. Instead of directly manipulating each qubit to perform universal quantum logic gates or measurements, ADQC uses a fixed two-qubit interaction to couple the memory register of a quantum computer to an ancilla qubit. By measuring the ancilla, the measurement-induced back-action on the system performs the desired logical operations.

By demanding that the ancilla-system qubit interaction should lead to unitary and stepwise deterministic evolution, and that it should be possible to standardise the computation, that is, applying all global operations at the beginning, we are able to place conditions on the interactions that can be used for ADQC. We prove there are only two such classes of interactions characterised in terms of the non-local part of the interaction operator. This leads to the definition of a new entanglement resource called \emph{twisted graph states} generated from  non-commuting operators. The ADQC model is  formalised in an algebraic framework similar to the Measurement Calculus \cite{Mcal06}. Furthermore, we present the notion of  \emph{causal flow} for twisted graph states, based on the stabiliser formalism, to characterise the determinism. Finally we demonstrate compositional embedding between ADQC and both the one-way and circuit models which will allow us to transfer recently developed theory and toolkits of measurement-based quantum computing directly into ADQC.
\end{abstract}


\maketitle
\section{Introduction}\label{intro}

There are two main paradigms which have driven both the theory and
implementation of quantum computation; gate-based quantum computing
(GBQC)~\cite{Deutsch89}, and measurement-based quantum computing
(MBQC)~\cite{RB01}. Though these two models are computationally equivalent,
in practice each has their own advantages and disadvantages which have major
implications for the choice of physical system, design, and operation. We
introduce a new paradigm called ancilla-driven quantum computing which
combines features of both models, in order to parallelise the architecture of quantum computers, to decrease decoherence effects, and simplify their physical implementation and operation. 

GBQC requires, in general, arbitrary networks of dynamic operations which
in turn complicates the design and characterisation of the entire
computer. There are many scenarios where it would be desirable to physically separate preparation, measurement, and coherent interaction regions to reduce
control complexity, circuitry congestion, and decoherence due to
cross-talk. In contrast, MBQC performs computation
purely through single-qubit measurement on a pre-existing static
multi-partite entangled state, distilling and processing non-local
correlation. However, the generation of the initial highly entangled
state, incorporation of quantum error correction and fault-tolerance,
and individual measurement of each qubit are issues in many candidate
systems.

Our new model of ancilla-driven quantum computing (ADQC) is partly inspired by the previous works of Andersson and Oi in~\cite{AO07}, and also Perdrix and Jorrand in~\cite{PJ04}. Andersson and Oi introduced an efficient method to implement any generalised quantum measurement by coupling the system with an ancilla qubit. The method, however, assumes arbitrary dynamic coupling operations between ancilla and system. Perdrix and Jorrand, on the other hand, describe a probabilistic version of MBQC in terms of a Turing machine where one can view the read-write head as an ancillary qubit, though this is not the way the model is presented. Moreover their approach still requires direct manipulation of the memory register and dynamic global measurement operators.

ADQC attempts to overcome such issues by performing computation where the memory register (input
data) can only be remotely manipulated through interaction with a
supply of prepared ancillas. In other words, instead of directly
manipulating data qubits to perform universal quantum logic gates or
measurements, ADQC uses a fixed two-qubit unitary interaction to
couple the memory register of a quantum computer to an ancilla qubit. 
By measuring
the ancilla, the measurement-induced back-action on the system
performs the desired logical operation. Practically, a single fixed
unitary interaction coupling the data and ancilla qubits greatly
simplifies design, construction, and operation of the computer since
only one particular discrete operation needs to be generated and
characterised. Furthermore, separating interaction and measurement
leads to a parallel structure with possibly reduced decoherence \cite{BK06}. A
requisite interaction for ADQC already exists in a variety
of physical systems ranging from ion micro-traps, neutral atoms,
nuclear spin donors in semiconductors, SQUIDs and cavity QED which
greatly increases the scope for implementation of the core ideas. ADQC
also naturally benefits from optimisation of the qubit species
employed for memory and ancilla. Memory qubits can be chosen for long
coherence time at the expense of being static and difficult to
manipulate directly, whilst ancilla qubits may be chosen for high
mobility and rapid initialisation and measurement, e.g. donor nuclear
spins in isotopically pure silicon as memory and electron spins
conveyed via charge transport by adiabatic passage as ancilla
in solid state quantum computing.

So far we have mentioned only the practical advantages of our proposed
architecture. The formalisation of the computational model underlying
ADQC, which is the focus of the current paper, leads to the
introduction of a new multi-partite entanglement resource. Only
recently has it been demonstrated that a very restricted class of
multi-partite entangled states are useful for universal MBQC~\cite{GFE08,BMW08}. However a full characterisation of such
states~\cite {GESP2007} remains an open problem which this paper aims
to make progress upon.

The entangled graph states~\cite{graphstates} have emerged as an elegant and
powerful quantum resource, especially for measurement-based quantum
computation (MBQC)~\cite{RB01}. Many important results on their entanglement
properties~\cite{NDMB07}, information flow~\cite{Flow06,g-flow},
implementation~\cite{BK05}, and novel applications in
cryptography~\cite{MS08,BFK08}, are due to their 
simple description. The generating operator for graph states, called
controlled-phase, is a symmetric and commuting operator which leads to a
simple graphical notation and hence the name for these states.  Additionally,
the elegant result by van den Nest \textit{et al.}~\cite{VDD04,NDMB07} shows
that any stabiliser state is equivalent to a graph state up to local Clifford
operators. This greatly expands the scope of the results for graph states, and leads to a
natural extension of the above constructions into stabiliser states, as well
as allowing a convenient graphical notation for a very general class of
states. If we consider \emph{open} graph states, graph states where some
nodes (called input nodes) are given in arbitrary states (rather than being
prepared in a particular fixed state which is the case for graph states)
much of the theory still follows. However open stabiliser states with
arbitrary input nodes no longer fulfil the pre-requisites of the theorem by
van den Nest \textit{et. al.}, and in general they do not admit a trivial
graphical notation.

Moreover, not all two-qubit interactions between system and ancillary qubits can be used for ADQC. To characterise interactions that will eventually enable universal ADQC we demand  that the ancilla-system interaction leads to a unitary and stepwise deterministic evolution of the system qubit, and that it should be possible to standardise the computation, that is, applying all global operations at the beginning. By doing so we are able to place conditions on the possible interactions resulting in two classes of interactions that are necessary and sufficient for ADQC. This naturally leads to the definition of a particular class of open stabiliser states,
called \emph{twisted graph states} which, despite being generated by
non-commuting operations, still admits a simple graph representation.
They form the key ingredient for ADQC. 
We then show how this new class of states can be viewed as open graph states up to somelocal swap operations. We also develop an algebraic framework similar to the
measurement calculus, which is the mathematical framework underlying
MBQC computation. This makes it possible to derive the standardisation theory for the ADQC patterns of computation. As we will see, any ADQC computation requires
a classical control structure to compensate for the probabilistic
nature of the measurement. In order to characterise the determinism, we introduce the notion of \emph{causal  flow} for twisted graph states based on the stabiliser formalism. Compared to the open graph state, the stabiliser state has a more complicated and global structure. 
We construct, however, direct translations between ADQC and MBQC for a 
subclass of deterministic patterns with flow which preserve depth, in order to prove that ADQC is as parallel as MBQC. We also present the embedding between GBQC and
ADQC and show how, as for MBQC, a separation in depth can be obtained.

\section{Preliminaries}\label{s-pre}

We briefly review the required concepts from quantum computing. A more
detailed introduction can be found in~\cite{NC00}. Let $\mcl H$ denote a
2-dimensional complex vector space, equipped with the standard inner
product. We pick an orthonormal basis for this space, label the two basis
vectors $\ket{0}$ and $\ket{1}$, and identify them with the
vectors $(1,0)^T$ and $(0,1)^T$, respectively.  A
\emph{qubit} is a unit length vector in this space, and so can be expressed
as a linear combination of the basis states:
$$
\alpha_0\ket{0}+\alpha_1\ket{1}=\left(\begin{array}{c}\alpha_0\\
\alpha_1 \end{array}\right).
$$
Here $\alpha_0,\alpha_1$ are complex \emph{amplitudes},
and $|\alpha_0|^2+|\alpha_1|^2=1$.

An \emph{$m$-qubit state} is a unit vector in the $m$-fold tensor space $\mcl
H\otimes\cdots\otimes \mcl H$.  The $2^m$ basis states of this space are the
$m$-fold tensor products of the states $\ket{0}$ and $\ket{1}$.  We
abbreviate $\ket{1}\otimes\ket{0}$ to $\ket{1}\ket{0}$ or $\ket{10}$. With
these basis states, an $m$-qubit state $\ket{\phi}$ is a $2^m$-dimensional
complex unit vector
$$
\ket{\phi}=\sum_{i\in\{0,1\}^m}\alpha_i\ket{i}.
$$
There exist quantum states that cannot be written as the tensor
product of other quantum states. Such states are called \emph{entangled} states, \eg~ $\ket {00} + \ket {11}$.

We use $\bra{\phi}=\ket{\phi}^*$ to denote the conjugate transpose of the
vector $\ket{\phi}$, and $\inp{\phi}{\psi}=\bra{\phi}\cdot\ket{\psi}$ for the
inner product between states $\ket{\phi}$ and $\ket{\psi}$.  These two states
are \emph{orthogonal} if $\inp{\phi}{\psi}=0$.  The \emph{norm} of
$\ket{\phi}$ is $\norm{\phi}=\sqrt{\inp{\phi}{\phi}}$.

A quantum state can evolve by a unitary operation or by a measurement. A
\emph{unitary} transformation is a linear mapping that preserves the norm of
the states. If we apply a unitary $U$ to a state $\ket{\phi}$, it evolves to
$U\ket{\phi}$. The \emph{Pauli operators} are a well-known set of unitary
transformations for quantum computing:
\begin{equation*} X= \begin{pmatrix} 0 & 1 \\
1 & 0
\end{pmatrix}, \,\,\, Y = \begin{pmatrix} 0 & -i \\ i & 0 \end{pmatrix}
, \,\,\, Z= \begin{pmatrix} 1& 0 \\0 & -1 \end{pmatrix}\, ,
\end{equation*}
and the \emph{Pauli group} on $n$ qubits is generated by Pauli
operators. Several other unitary transformations that we will use in this
paper are the identity $\mathbbm{1}$, the \emph{phase} gate $P(\alpha)$, of which
$P(\pi/4)$ and $P(\pi/2)$ are a special cases, the Hadamard $H$, the
controlled-$Z$ ($\ctR Z$) and the SWAP operation:

\AR{
  \mathbbm{1} := \begin{pmatrix}
 1 & 0 \\
 0 & 1
\end{pmatrix}, \;\;\;\;
P(\alpha) := \begin{pmatrix}
  1 & 0 \\
  0 & e^{i\alpha}
\end{pmatrix},
\;\;\;
H :=\frac{1}{\sqrt{2}}
\begin{pmatrix}
  1 & 1 \\
  1 & -1
\end{pmatrix},
\\\\
\ctR Z :=
\begin{pmatrix}
    1 & 0 & 0 & 0   \\
    0 & 1 & 0 & 0 \\
    0 & 0 & 1 & 0 \\
    0 & 0 & 0 & -1
\end{pmatrix}, \;\;\;\;
\text{SWAP} :=
\begin{pmatrix}
    1 & 0 & 0 & 0   \\
    0 & 0 & 1 & 0 \\
    0 & 1 & 0 & 0 \\
    0 & 0 & 0 & 1
\end{pmatrix}.
}
The \emph{Clifford group} on $n$ qubits is generated by the matrices $Z$,
$H$, $P({\pi/2})$ and $\ctR Z$, and is the normaliser of the Pauli group. This
set of matrices is not universal for quantum computation, but by adding any
single-qubit gate not in the Clifford group (such as $P({\pi/4})$), we do get
a set that is approximately universal for quantum computing \cite{NC00}.

The most general measurement allowed by quantum mechanics is specified by a
family of positive semi-definite operators $E_i=M_i^*M_i$, $1\leq i\leq k$,
subject to the condition that $\sum_i E_i= \mathbbm{1}$. A projective measurement is
defined in the special case where the operators $E_i$ are projections. Let
$\ket{\phi}$ be an $m$-qubit state and $\mcl
B=\{\ket{b_1},\ldots,\ket{b_{2^m}}\}$ an orthonormal basis of the $m$-qubit
space. A projective measurement of the state $\ket{\phi}$ in the $\mcl B$
basis means that we apply the projection operators $P_i=\ket{b_i}\bra{b_i}$
to $\ket{\phi}$.  The resulting quantum state is $\ket{b_i}$ with probability
$p_i=|\inp{\phi}{b_i}|^2$. An important class of projective measurements are
Pauli measurements, \ie~projections onto eigenstates of Pauli operators.

So far we have dealt with \emph{pure} quantum states. A more general
representation with density matrices also allows us to describe open
physical systems, where one can prepare a classical stochastic mixture
of pure quantum states, called \emph{mixed} quantum states. For a
system in a pure state $\ket{\psi}$, the density matrix is just the
projection operator $\proj{\psi}$. Suppose that we only know that a
system is one of several possible states
$\ket{\psi_1},\ldots,\ket{\psi_k}$ with probabilities $p_1,\ldots,p_k$
respectively.  We define the density matrix for such a state to be
\[ \rho = \sum_{i=1}^k p_i\proj{\psi_i}.\] The most general physical
operator that acts over density matrices is a completely positive
trace preserving map (CPTP)
$\mathcal{E}:\mathcal{B}(\hilb_1)\to\mathcal{B}(\hilb_2)$ with Kraus
decomposition
\[ \mathcal{E}(\rho) = \sum_m \KO_{m}\rho\adj{\KO_{m}} \] where the
$\KO_m:\hilb_1\to\hilb_2$, $\mathcal{B}(\hilb)$ is the Banach space of bounded
linear operators and we require that
\[ \sum_m \adj{\KO_m} \KO_m = \mathbbm{1}.\]

Any unitary operation $U$ can be approximated with a circuit~$C$, using gates
from a fixed universal set of gates. The \emph{size} of a circuit is the number of gates and
its \emph{depth} is the largest number of gates on any input-output
path. Equivalently, the depth is the number of layers that are required for
the parallel execution of the circuit, where a qubit can be involved in at
most one interaction per layer. In this paper, we adopt the model according
to which at any given time-step, a single qubit can be involved in at most one
interaction. This differs from the \emph{concurrency} viewpoint, according to
which all interactions for commuting operations can be done simultaneously.

\subsection{Measurement-based model}\label{ss-mbqc}

We give a brief introduction to measurement-based quantum
computing (MBQC)~\cite{RB01,RB02,RBB03}. A more detailed description
is available in~\cite{Jozsa05,Nielsen05,BB06,Mcal06}. Our notation
follows that of~\cite{Mcal06}. In MBQC, computations are represented
as \emph{patterns}, which are sequences of \emph{commands} acting on
the qubits in the pattern. These commands are of four types:
\begin{enumerate}
\item $N_i$ is a one-qubit preparation command which prepares the auxiliary
  qubit~$i$ in state $\ket{+}= \frac{1}{\sqrt{2}} (\ket{0} + \ket{1})$. The
  preparation commands can be implicit from the pattern: when not specified,
  all non-input qubits are prepared in the $\ket{+}$ state.
\item $\et ij$ is a two-qubit entanglement command which applies the
  controlled-$Z$ operation, $\ctR Z$, to qubits $i$ and $j$. Note that the
  $\ctR Z$ operation is symmetric so that $E_{ij} = E_{ji}$. Also, $E_{ij}$
  commutes with~$E_{jk}$ and thus the ordering of the entanglement commands in
  not important.
\item $\M\al i$ is a one-qubit measurement on qubit~$i$ which depends
  on parameter $\al \in [0,2\pi)$ called the \emph{angle of
    measurement}. $\M\al i$ is the orthogonal projection onto the states
 \begin{align*}
 \ket{+_\al}&=\ost(\ket0+ e^{i\al}\ket1)\\
\ket{-_\al}&=\ost(\ket0-\ei\al\ket1),
\end{align*}
followed by a trace-out operator, since measurements are destructive. We
denote the classical outcome of a measurement performed at qubit $i$ by
$m_i\in\ztwo$. We take the specific convention that $m_i=0$ if the
measurement outcome is~$\ket{+_\al}$, and that~$m_i=1$ if the measurement
outcome is~$\ket{-_\al}$. Outcomes can be summed together resulting in
expressions of the form
\begin{equation*}
m=\sum_{i\in I} m_i
\end{equation*}
which are called \emph{signals}, and where the summation is understood
as being done modulo~$2$.  The \emph{domain} of a signal is the set of
qubits on which it depends (in this example, the domain of~$m$
is~$I$).
\item $X_i$ and $Z_i$ are one-qubit Pauli corrections which
correspond to the application of the Pauli $X$ and $Z$ matrices,
respectively,  on qubit $i$.
\end{enumerate}

In order to obtain universality, we have to add a classical control
mechanism, called \emph{feed-forward}, which allows measurement angles
and corrections to be dependent on the results of previous
measurements~\cite{RB01,Mcal06}. Let~$m$ and~$n$ be signals. Dependent
corrections are written as $\cx im$ and $\cz in$ and dependent
measurements are written as $\mLR \al imn $.  The meaning of
dependencies for corrections is straightforward: $\cx i0=\cz i0= \mathbbm{1}$ (no
correction is applied), while $\cx i1=\Cx i$ and $\cz i1=\Cz i$\,.  In
the case of dependent measurements, the measurement angle depends on
$m$, $n$ and $\al$ as follows:
\begin{align} \label{msem}
\mLR{\al} imn = \m{(-1)^m\al+n\pi} i
\end{align}
so that, depending on the parity of~$m$ and $n$, one may have to
modify the angle of measurement~$\al$ to one of $-\al$, $\al+\pi$ and
$-\al+\pi$. These modifications correspond to conjugations of
measurements under $X$ and $Z$:
\begin{align} \label{e-Xact}
\cx im\M{{\al}}i \cx im&=\M{{(-1)^m\al}}i\\
\label{e-Zact} \cz in\M{{\al}}i \cz in&=\M{{\al+n\pi}}i
\end{align}
and we will therefore refer to them as the $X$- and $Z$-actions, or
alternatively as the $X$- and $Z$-dependencies. Since measurements are
destructive, the above equations simplify to
\begin{align}
  \M{{\al}}i \cx im&=\M{{(-1)^m\al}}i\label{xmx}\\
  \M{{\al}}i \cz in&=\M{{\al+n\pi}}i\label{zmz}.
\end{align}
Note that these two actions commute, since $-\al+\pi=-\al-\pi$
up to $2\pi$, and hence the order in which one applies them does not
matter.

A \emph{pattern} is defined by the  choice of a finite set $V$ of
qubits, two not necessarily disjoint sets $I \subseteq V$ and $O
\subseteq V$ determining the pattern inputs and outputs, and a
finite sequence of commands acting on $V$. We require that  no
command depend on an outcome not yet measured, that no command act
on a qubit already measured, that a qubit be measured if and only if
it is not an output qubit and that a qubit be prepared if and only if
it is not an input qubit. This set of rules is known as the
\emph{definiteness} condition.

A pattern is said to be in \emph{standard form} if all the preparation commands
$N_i$ and entanglement operators $\et ij$ appear first in its command
sequence, followed by measurements and finally corrections. A pattern
that is not in standard form is called a \emph{wild pattern}. Any wild
pattern can be put in its unique standard form~\cite{Mcal06}; this
form can reveal implicit parallelism in the computation. The procedure
of rewriting a pattern in its standard form is called
\emph{standardisation}. This can be done by applying the following
rewrite rules:
 \EQ{ \label{e-EX}
\et ij\cx im&\Rar&\cx im\cz jm\et ij \\
\label{e-EZ}
\et ij\cz im&\Rar&\cz im\et ij\\
\label{e-MX}
\mLR\al imn\cx i{p}&\Rar&\mLR\al i{m+p}{n}\\
\label{e-MZ} \mLR{\al} imn\cz i{p}&\Rar&\mLR\al i{m}{n+p}. }
The rewrite rules also contain the following
\emph{free commutation rules} which tell us that if we are dealing
with disjoint sets of target qubits, then measurement, corrections and
entanglement commands commute pairwise \cite{Mcal06}, so that
\begin{align}
\et ij\CO{\vec k}&\Rar\CO{\vec k}\et ij \quad\hbox{where $A$ is not an
  entanglement command}\\
\CO{\vec k}\cx im&\Rar\cx im\CO{\vec k} \quad\hbox{where $A$ is not
a correction command}\\
\CO{\vec k}\cz im&\Rar\cz im\CO{\vec k} \quad\hbox{where $A$ is not a
  correction command}
\end{align}
where~$\vec k$ represent the qubits acted upon by command~$A$, and are
distinct from~$i$ and~$j$. 

Standardisation allows us to graphically present the global operation of a pattern. We define an \emph{open graph state} $(G,I,O)$ to consist of an undirected graph $G$
together with two subsets of nodes $I$ and $O$, called inputs and
outputs. We write $V$ for the set of vertices in $G$, $E$ for the set
of edges, $I^c$, and $O^c$ for the complements of $I$ and $O$ in $V$
and $E_G:=\prod_{\{i,j\}\in E}E_{ij}$ for the global entanglement
operator associated with $G$. Trivially, any standard pattern has a
unique underlying open graph state, obtained by neglecting 
measurements and correction commands.

We now give a condition on geometries under which it is possible to
synthesize a set of dependent corrections such that the obtained
pattern is uniformly and strongly deterministic, \ie~all the
branches of the computation are equal, independently of the angles of
the measurements. In what follows, $x \sim y$ denotes that  $x$ is adjacent
to $y$ in $G$ and $N_{I^c}$ denotes the sequence of preparation
commands $\prod_{i\in I^c} N_i$.

\begin{definition} A \emph{flow} $(f,\preceq)$ for a geometry $(G,I,O)$ consists of
a map $f:O^c\rar I^c$ and a partial order $\preceq$ over $V$ such
that for all $x\in O^c$:
\begin{itemize}
\item (i)~~$x \sim f(x)$;
\item (ii)~~$x \preceq f(x)$;
\item (iii)~~for all $y \sim f(x)$,  $x \preceq y$\,.
\end{itemize}
\end{definition}

The coarsest order $\preceq$ for which~$(f, \preceq)$ is a flow is
called the \emph{dependency order} induced by~$f$ and its depth is called \emph{flow depth}.

\begin{theorem} \label{t-oldflow} Suppose the geometry $(G,I,O)$ has
flow~$f$. Then the pattern
 \AR{ \mcl
P_{f,G,\vec\al}&:=&\prod\limits_{i\in O^c}{\!\!\!}^{\preceq}\,\,\,
\Big(\cx{f(i)}{s_i}\prod\limits_{\substack{k \sim f(i) \\ k \ne
i}}\cz{k}{s_i}\m{\al_i}i\Big) E_G } where the product follows the
dependency order $\preceq $ of $f$,  is uniformly and strongly
deterministic, and realizes the unitary embedding
 \AR{
U_{G,I,O,\vec\al}&:=& {2}^{|O^c|/2}\;\Big(\prod\limits_{i\in
O^c}\bra{{+_{\al_i}}}_i \;\Big)\, E_G. }
\end{theorem}

If the underlying geometry of a pattern has a flow and its pattern command
sequence is constructed as given by the flow theorem, we call
this pattern a \emph{pattern with flow}.

\section{Ancilla-Driven Model}\label{s-model}

As mentioned in the introduction, in ancilla-driven quantum computing we are interested in  two essential properties:
\begin{itemize}
\item The only global operation is a fixed two-qubit interaction between
  any ancilla and system qubit.
\item Only ancilla qubits will be measured.
\end{itemize}

We introduce the ADQC model within an algebraic framework similar to 
that of the measurement calculus recalled in the previous section. We
have a set of fixed basic commands described below where the indices
$i$, $j$, $\dots$ represent the qubits on which each of these
operations apply. A \emph{pattern} is a sequence of commands defined
over a set of qubits in the list $V$, called \emph{computation space}, where the particular sub-list $S$ represents the \emph{system} qubits (we may refer to them as data or
memory register) and the rest $A=V\setminus S$ are the \emph{ancilla}
qubits. We define an arbitrary pure single qubit state
by \AR{ \ket{+_{\ta,\phi}}=\cos(\frac{\ta}{2})\ket 0 +
  e^{i\phi}\sin(\frac{\ta}{2})\ket 1 \, , } and denote its orthogonal
state (the opposite point in the Bloch sphere) with \AR{
  \ket{-_{\ta,\phi}}=\sin(\frac{\ta}{2})\ket 0 -
  e^{i\phi}\cos(\frac{\ta}{2})\ket 1\,, } where $0\leq \ta \leq \pi$
and $0\leq \phi \leq 2\pi$.

\begin{itemize}
\item {\bf Preparation.} $N^{\ket \psi}_a$ ($a \in A$) prepares an
  ancilla qubit in the state $\ket \psi$.
\item {\bf Interaction.} $\etil a s$ ($s \in S, a \in A$) entangle a
  system qubit and an ancilla qubit with interaction operator
  $\ctR Z$ followed by Hadamard  on each qubit: $$\widetilde{\ctR
    Z}:= H_s\otimes H_a\ctR Z_{as}$$
    Note that a $\ctRZSWAP$ interaction would be another possible choice. In section \ref{sec:characteriseE} we will prove that $\ctR Z$ and $\ctRZSWAP$ are the only two possible interactions, up to local unitary equivalence, that allow for universal ADQC. Choosing one over the other will depend on the natural dynamics of the physical system used for an implementation.

 \item {\bf Ancilla Measurement.} $\GM \lambda \alpha a$ ($a \in A$)
  measures qubit $a$ on plane $\lambda\in \{\cplane XY,\cplane XZ,
  \cplane YZ\}$, defined by orthogonal projections into:
\begin{itemize}

\item $\ket {\pm_{\cplane XY,\alpha}}:=\ket{\pm_{\frac \pi 2,\alpha}}$
  if $\lambda = \cplane XY$
\item $\ket {\pm_{\cplane XZ,\alpha}}:=\ket{\pm_{\alpha,0}}$ if $\lambda = \cplane XZ$
\item $\ket {\pm_{\cplane YZ,\alpha}}:=\ket{\pm_{\alpha,\frac \pi 2}}$
  if $\lambda = \cplane YZ$
\end{itemize}
with the convention that $\ket{+_{\ta,\phi}}\bra{+_{\ta,\phi}}_a$
corresponds to the outcome $m_a=0$, while
$\ket{-_{\ta,\phi}}\bra{-_{\ta,\phi}}_a$ corresponds to $m_a=1$. The
propagation of dependent corrections (next command) defines dependent
measurement: \AR{ \GMLR \lambda \alpha a m n := \GM \lambda
  \alpha a X_{a}^{m} Z_{a}^{n} } where $m, n, \dots$ are module 2
summation of several measurements outcomes, also called \emph{signals}. The \emph{domain} of a signal is the set of
qubits on which it depends.\footnote{Depending on the context we sometimes use the notation $m$ for syntax, i.e., a set of qubits (representing a formal sum) and sometimes for semantics, i.e., 0 or 1.}

\item {\bf Corrections.} $X_i$ and $Z_i$ ($i \in V$), 1-qubit Pauli
  operators. As in MBQC, to control the non-determinism of the
  measurement outcomes certain local corrections will depend upon
  previous measurement outcomes. These will be written as $C_i^{m}$,
  with $C_i^0= \mathbbm{1}$, and $C_i^1=C_i$.
\end{itemize}

We write $\hil V$ for the associated quantum state space $\otimes_{i\in V}\ctwo$. To run
a pattern, one prepares the system qubits in some given input state
$\Psi\in\hil S$, while the ancilla qubits are all prepared according
to the $N$ commands in fixed $\ket{\psi}$ states. The commands
are then executed in sequence, and finally the result of the pattern
computation is read back from the system qubits\footnote{Preparation
  and readout of the system qubits can be performed by using suitable
  ancilla states and measurements.}. Similar to the one-way model we will consider only patterns satisfying definiteness conditions. 

The main differences between ADQC and MBQC are: (1) The interaction operator being $\widetilde{\ctR Z}$ instead of $\ctR Z$, which still belongs to the so-called Clifford group, the normaliser of the Pauli group; (2) Only ancilla qubits can be measured, that is to say in the terminology of MBQC any ADQC pattern has the same number of inputs and outputs which are overlapping (system qubits). Apart from universality, which we will prove later most of the theory of measurement calculus~\cite{Mcal06} which was developed for the one-way quantum computer can be easily adapted to ADQC. For completeness we briefly review this here.

The first way to combine patterns is by composing them. Two patterns
$\mfr P_{1}$ and $\mfr P_{2}$ may be composed if $S_1=S_2$. Provided
that $\mfr P_1$ has as many system qubits as $\mfr P_2$, by renaming
these qubits, one can always make them composable. However it is
important to emphasise that since the $\etil ij$ operators are
non-commuting, their order of appearance in each pattern must be
preserved under the renaming and composition. The other way of
combining patterns is to tensor them.  Two patterns $\mfr P_{1}$ and
$\mfr P_{2}$ may be tensored if $V_1\cap V_2=\emptyset$.  Again one
can always meet this condition by renaming qubits in a way that these
sets are made disjoint.

\subsection{The semantics of patterns}

We present a formal operational semantics for ADQC patterns as a
probabilistic labelled transition system, similar
to~\cite{Mcal06}. Besides quantum states, one needs a classical state
recording the outcomes of the successive measurements made in a
pattern.  If we let $U$ stand for the finite set of qubits that are
still active (i.e.\ not yet measured) and $W$ stands for the set of
qubits that have been measured (i.e.\ they are now just classical bits
recording the measurement outcomes), it is natural to define the
computation state space as
 \AR{ \mcl C&:=&\Sigma_{U,W} \hil U\times\ztwo^W.  } 
 In other words, the computation states form a
$U,W$-indexed family of pairs $q$, $\Ga$, where $q$ is a quantum state
from $\hil U$ and $\Ga$ is a map from some $W$ to the outcome space
$\ztwo$.  We call this classical component $\Ga$ an \emph{outcome
  map}, and denote by $\emptyset$ the empty outcome map in
$\ztwo^\emptyset$. We need further notation.  For any
signal $m$ and classical state $\Ga\in\ztwo^W$, such that the domain
of $m$ is included in $W$, we take $m_\Ga$ to be the value of $m$
given by the outcome map $\Ga$.  That is to say, if $m=\sum_I m_i$,
then $m_\Ga:=\sum_I\Ga(i)$ where the sum is taken in $\ztwo$.  Also if
$\Ga\in\ztwo^W$, and $x\in\ztwo$, we define
\AR{\Ga[x/i](i)=x,\,\Ga[x/i](j)=\Ga(j)\hbox{ for }j\neq i} which is a
map in $\ztwo^{W\cup\ens i}$.

We may now view each of our commands as acting on the state space
$\mcl C$: \AR{
  q,\Ga&\slar{N_i^{\ket \psi}}&q\otimes\ket{\psi}_i,\Ga\\
  q,\Ga&\slar{\etil ij}&\widetilde{\ctR Z}_{ij} q,\Ga\\
  q,\Ga&\slar{\cx im}&\cx i{m_\Ga} q,\Ga\\
  q,\Ga&\slar{\cz im}&\cz i{m_\Ga} q,\Ga\\
  U\cup\{i\},W,q,\Ga&\slar{\MS{\lambda,\al}imn}&U,W\cup\{i\},{\oqbb{\lambda,\al_\Ga}}_iq,\Ga[0/i]\\
  U\cup\{i\},W,q,\Ga&\slar{\MS{\lambda,\al}imn}&U,W\cup\{i\},{\oqbnb{\lambda,\al_\Ga}}_iq,\Ga[1/i]
} where $\al_\Ga=(-1)^{m_\Ga}\al+n_\Ga\pi$.  We introduce an
additional command called \emph{signal shifting}: \AR{ q,\Ga&\slar{\ss
    i{m_\Ga}}&q,\Ga[\Ga(i)+m_\Ga/i] } It consists in shifting the
measurement outcome at $i$ by the amount $m_\Ga$.  Note that the
$Z$-action leaves measurements globally invariant, in the sense that
$\oqb{\al+\pi},\oqbn{\al+\pi}=\oqbn{\al},\oqb{\al}$.  Thus changing
$\al$ to $\al+\pi$ amounts to exchanging the outcomes of the
measurements, and one has
 \EQ{\MS\al i{m_\Ga}{n_\Ga}&=&\ss i{n_\Ga}\,\MS\al
  i{m_\Ga}0\label{split}.}
Signal shifting allows us to dispose of the $Z$
action of a measurement, sometimes resulting in convenient
optimisations of standard forms. In the rest of the paper, for simplicity, we omit the superscript $\Ga$ on the measurement outcomes. 

The convention is that when one does a measurement the resulting state is
\emph{renormalised} and the probabilities are associated with the
transition.  We do not adhere to this convention here, but instead leave the
states unnormalized.  The reason for this choice is that in this
way, the probability of reaching a given state can be read off its norm,
and the overall treatment is simpler.  

\subsection{Denotational Semantics}

We now present the denotational semantics of ADQC patterns. If $n$ is
the number of measurements then the run may follow $2^n$ different
branches.  Each branch is associated with a unique binary string
$\mathbf{n}$ of length $n$, representing the classical outcomes of the
measurements along that branch, and a unique \emph{branch map} 
(Kraus operator) $\KO_{\mathbf{n}}$ representing the linear transformation from $\hil S$
to $\hil S$ along that branch. This map is obtained from the (un-normalised) operational semantics via the sequence
$(q_i,\Ga_i)$ with $1\leq i\leq m$ (where $m$ is the total number of commands), such that 
\AR{
q_1,\Ga_1=q\otimes\ket{\hskip-.4ex+\ldots+},\emptyset\\
\hbox{and for all }i\leq m:q_{i-1},\Ga_{i-1}\slar{\KO_i}q_{i},\Ga_{i}.
}
and all measurement commands in the sequence $\{\KO_i\}$ have been replaced by appropriate projections corresponding to the outcome index $\mathbf{n}$.

\begin{definition} A pattern $\mfr P$ realizes a map
on density matrices $\rho$ given by $\rho\mapsto
\sum_{\mathbf{s}}\KO_{\mathbf{s}}(\rho) \KO_{\mathbf{s}}^{\dag}$.  We
write $\sem{\mfr P}$ for the map realised by $\mfr P$.  \end{definition} It is then
easy to prove~\cite{Mcal06} that each pattern realizes a completely
positive trace preserving (CPTP) map and if a pattern is strongly
deterministic (see section 2), then it
realizes a unitary embedding~\cite{Mcal06}. Hence the denotational
semantics of a pattern is a CPTP-map. It is also compositional, as the following theorem shows.

\begin{theorem} For two patterns $\mfr{P}_1$ and $\mfr{P}_2$ we have
$\sem{\mfr{P}_1\mfr{P}_2} = \sem{\mfr{P}_2}\sem{\mfr{P}_1}$ and
$\sem{\mfr{P}_1\otimes\mfr{P}_2} =
\sem{\mfr{P}_2}\otimes\sem{\mfr{P}_1}.$ \end{theorem} 

\noindent\textbf{Proof.}  Recall
that two patterns $\mfr P_1$, $\mfr P_2$ may be combined by
composition provided $\mfr P_1$ has as many system qubits as $\mfr
P_2$. Suppose this is the case, and suppose further that $\mfr P_1$
and $\mfr P_2$ respectively realise some CPTP-maps $T_1$ and $T_2$.
We need to show that the composite pattern $\mfr P_2\mfr P_1$ realizes
$T_2T_1$. Indeed, the two diagrams representing branches in $\mfr P_1$
and $\mfr P_2$:

\vskip0.1cm
{\footnotesize
\AR{
\xymatrix@=10pt@M=3pt@R=20pt@C=7pt{
{}\hil {S_1}\ar[d]\ar@{.>}[rr]
&&
{}\hil {S_1}\ar@{=}
&
{}\hil {S_2}\ar[d]\ar@{.>}[rr]
&&
{}\hil {S_2}
\\
{}\hil {S_1}\times\ztwo^{\emptyset}\ar[r]^{p_1}&
{}\hil {V_1}\times\ztwo^{\emptyset}\ar[r]^{}&
{}\hil {S_1}\times\ztwo^{V_1\setminus S_1}\ar[u]
&
{}\hil {S_2}\times\ztwo^{\emptyset}\ar[r]^{p_2}&
{}\hil {V_2}\times\ztwo^{\emptyset}\ar[r]^{}&
{}\hil {S_2}\times\ztwo^{V_2\setminus S_2}\ar[u]
}
}
}
\vskip0.1cm

\noindent can be pasted together, since $S_1=S_2$, and $\hil
{S_1}=\hil {S_2}$.  But then it is enough to notice 1) that
preparation steps $p_2$ in $\mfr P_2$ commute with all actions in
$\mfr P_1$ since they are appled on disjoint sets of qubits, and 2) that no
action taken in $\mfr P_2$ depends on the measurements outcomes in
$\mfr P_1$.  It follows that the pasted diagram describes the same
branches as does the one associated to the composite $\mfr P_2\mfr
P_1$. A similar argument applies to the case of a tensor combination,
and one has that $\mfr P_2\otimes\mfr P_1$ realizes $T_2\otimes T_1$.
\qed \vskip 0.2cm

\subsection{Generating patterns}
\label{sec:universal}

In order to prove the universality we present two simple generic
patterns where only measurements in the $\cplane XY$ plane  ($M^{\al}$), Pauli
$Z$ measurements ($M^Z$) and preparations of the ancilla in the state $\ket +$ ($N$) are
sufficient. Note that a Pauli $Z$ measurement can be considered as a special case of a measurement in the $\cplane XZ$ or $\cplane YZ$ plane, with $\al = 0$.

The following one-parameter family $J(\al)$ generates all single-qubit
unitary operators \cite{generator04}: \AR{
  J(\al):=\ost\MA{1&\ei\al\\1&-\ei\al} } as any unitary operator $U$
on $\ctwo$ can be written: \AR{ U=e^{i\al}J(0)J(\ba)J(\ga)J(\da) } for
some $\al$, $\ba$, $\ga$ and $\da$ in $\mbb R$. Recall that the MBQC
implementation of the $J$ generator is the following pattern: \EQ{\label{e-Jpat1} \mfr
  J(-\al)&:=&\cx a{m_1}\GM {\cplane XY}{\al} s\et sa } 
where $s$ is the system qubit input, $a$ is the ancilla and $\et sa$ is
$\ctR Z$ operator \cite{generator04}. On the other hand, the following pattern also implements the $J$ gate, but now the ancilla qubit $a$ will instead be measured:
 \EQ{\label{e-Jpat2} \mfr
  J(-\al)&:=&H_s\cz s{m_a}\GM {\cplane YZ} {\al} a\et sa, } where $H_s$
represents the application of a Hadamard gate on the system qubit. We
can now manipulate the new pattern to derive a generating pattern for the operator
$J$ in our model. We want to use  $\etil s a$ as the interaction command:  
\EQ{\nonumber \mfr
  J(-\al) &:=&H_s\cz s{m_a}\GM {\cplane YZ} {\al} a\et sa \\ \nonumber
  &=&\cx s{m_a}\GM {\cplane YZ} {\al} a H_s\et sa \\ \nonumber &=&\cx
  s{m_a}\GM {\cplane XY} {\al} a H_aH_s\et sa \\\label{e-Jpat3} &=&\cx
  s{m_a}\GM {\cplane XY} {\al} a \etil sa .} 
In addition to this, we only need a
generator for a two-qubit unitary such as $\ctR Z$ to obtain the
full universality. The MBQC pattern for controlled-$Z_{ij}$ ($\et i
j$ with $i,j \in S$) is, however, not desirable, as it is an operator between two
qubits of the system rather than an interaction between system and
ancilla qubits. Therefore the natural choice instead is to consider
the interaction  $\etil a{s^\prime} \etil as$. It is easy to
check that this, combined with Pauli $Z$ measurement of the ancilla, 
will give us a simple generating pattern for the two qubit operator $\widetilde{\ctR
  Z}$: 
  \EQ{\label{e-ctrZpat} \mfr {\widetilde{\ctR Z}} &:=& \cx
  s {m_a} M^{Z}_a \etil a{s^\prime}\etil as .} 

Any unitary can then be
simulated by sequential and parallel compositions of the above
generating patterns, where the composition simply glues given patterns
over the common system qubits, while preserving the initial orders of
the commands. We will return to the important issue of how to
represent the composed pattern graphically, but in order to do so, we first have
to address the important feature of standardization in the ADQC model. That is, the
standardisation procedure which permits us to rewrite any well
defined patterns, e.g. obtained from composition, in a
standard form where all the preparation commands are applied first,
followed by the entangling, measurement, and finally correction
commands.

For simplicity, in the remainder of this paper we will restrict ourselves to a special class of patterns, namely, those using only ancillas of degree 1 with arbitrary $\cplane XY$  plane measurement and of degree 2 with Pauli $Z$ measurement.  However, the whole theory developed in this paper can easily be extended to the more general setting.

\subsection{Standardisation} \label{sec:adqc-standard}

Similar to the one-way model \cite{Mcal06} we present a simple calculus of local equations by which any general pattern can be put into a standard form where entanglement is
done first, then measurements, then corrections. The consequences of the existence of such a procedure (called standardisation) are far-reaching and is explained in details in \cite{Mcal06}. We just recall that since entangling comes first, one never has to do ``on
the fly'' entanglements and the rewriting of a pattern to
standard form reveals parallelism in the pattern computation.  In a general
pattern, one is forced to compute sequentially and to strictly obey the
command sequence, whereas, after standardization, the dependency structure
is relaxed, resulting in lower computational depth complexity.  It is known that any MBQC model can admit a standardisation procedure if and only if the entangling command belongs to the normaliser group of the group generated by the correction commands~\cite{g-flow}. This
is the case for our ADQC model. The required rewrite rules are \EQ{
  \etil i j \cx is &=&\cx js\cz is\etil ij\label{etil-cx}\\
  \etil i j \cz is &=&\cx is\etil ij\label{etil-cz} .} 
  The rules for
propagation of the correction through measurement are the same as for 
MBQC, but with additional rules for the $M^Z$ measurement: \EQ{
  \mLR \al a m n\cx a p&=&\mLR \al a{m+p}{n}\label{mtil-x}\\
  \mLR \al a m n\cz a p&=&\mLR\al a {m}{n+p}\label{mtil-z}\\\nonumber\\
  M^Z_a \cx a m &=& \ss a m M^Z_a \label{mtil-zx}\\
  M^Z_a \cz a m &=& M^Z_a \label{mtil-zz} .}
We also have the same free commutation rewrite rules:
\begin{align}
\label{freecomm1}\etil ij\CO{\vec k}&\Rar\CO{\vec k}\etil ij \quad\hbox{where $A$ is not an
  entanglement command}\\
\CO{\vec k}\cx im&\Rar\cx im\CO{\vec k} \quad\hbox{where $A$ is not
  a correction command}\\
\label{freecomm3}\CO{\vec k}\cz im&\Rar\cz im\CO{\vec k} \quad\hbox{where $A$ is not a
  correction command}
\end{align}
where $\vec k$ represent the qubits acted upon by command $A$, and are
distinct from $i$ and~$j$.

Recall that the effect of a $Z$ correction on a qubit $a$ simply flips
the outcome of a measurement to be made on that qubit. Hence we can
replace the dependencies induced by the $Z$ correction by appropriate
operations over the measurement outcomes as described below. In what
follows, $m[n/m_i]$ denotes the
substitution of $m_i$ with $n$ in $m$, where $m$, $n$ are modulo 2
summations of several measurement outcomes:  \EQ{
  \mLR\al a m n &=&\ss a n\mR \al a m \\
  \cx j m \ss in &=& \ss i n \cx j {m[n+m_i/m_i]}\\
  \cz j m\ss in&=& \ss in \cz j{m[n+m_i/m_i]}\\
  \mLR \al j m n\ss ip&=&\ss ip\, \mLR \al j{m[p+m_i/m_i]}{n[p+m_i/m_i]}\\
  \ss i m \ss j n&=&\ss j n \ss i{m[n+m_j/m_j]}. } One can then use the
exact same method as for MBQC in order to prove that this rewrite
system has the desired properties of confluence and termination.

We emphasise  again that a key difference between ADQC and MBQC is the interaction command $\etil i j $ versus $\et i j$.  The explicit inclusion of the additional local Hadamard operations, $\etil i j = H_i H_j \et i j$, is necessary for universality. This is due to the fact that no system qubit can be directly measured but instead any operation has to be implemented via the ancilla. It is apparent that while rotations in the $z$-basis can be performed with only $\et i j = \ctR Z_{ij}$, no \emph{basis change} at the system can be implemented via the ancilla.

As shown in section 2, for any standard pattern in MBQC we can write its underlying
open graph state with qubits representing the nodes and $\et i j$ the
edges of the graph. Then, remarkably, only from the geometry of this
graph we can obtain the dependency structures to guarantee a
deterministic computation in MBQC. In other words, the simple graph
representation for the global operation defining pattern allows one to
determine dynamic properties directly from the static structure. Can
we still obtain similar properties for our new model? Despite the
non-commutativity of $\etil i j$ the answer is yes. In section 5 we define the twisted graph state which is the underlying
geometry of a given ancilla-driven pattern obtained from
standardisation and we present how one can directly construct the
dependency structure from their geometry.

\section{Characterisation of Interaction} \label{sec:characteriseE}

A central question in the theory of measurement-based quantum computing is the characterisation of the universal resources \cite{NDMB07, stabilisers, GESP2007, CTNstates}. Unlike these studies that analyse the computational power of a uniform class of multi-partite entangled states, we take a ``bottom up'' approach and identify basic \emph{building blocks} that can be composed to perform universal ADQC. This requires the characterisation of all two-qubit interactions which couple any two ancilla and system qubits, while satisfying certain desirable conditions such as \emph{stepwise determinism, unitarity and standardisation} that will lead to universality, in the `universal state preparation' sense \cite{NDMB07, stabilisers}. We obtain a full characterisation of universal resources for ADQC for these conditions, while in contrast, such a general result is not available for MBQC.

In this paper we will focus on stepwise determinism, that is a pattern which is deterministic after performing each single measurement together with all the Pauli corrections depending on the result of that measurement. Other computation strategies could be considered that are not stepwise deterministic, such as the (finitely) repeated application of the same operation in the scheme \cite{GESP2007, CTNstates}. Generalising ADQC to adapt to these less restricted strategies remains an open problem. 

We introduce the canonical decomposition of two-qubit unitaries \cite{ZVSW2003},  
\begin{equation} \label{eq:decomp}
	 \etil a s =(W'_a \otimes W_s) \, D_{as}  \, (V'_a \otimes V_s),
\end{equation}
where $W'_{a}, W_{s}$ and $V'_{a},V_{s}$ are single qubit unitaries and the diagonal global operation is
\begin{eqnarray}\label{e-Dformula}
	D_{as} (\alpha_x, \alpha_y, \alpha_z) = e^{-i (\alpha_x X_a \otimes X_s + \alpha_y Y_a \otimes Y_s + \alpha_z Z_a \otimes Z_s)}.
\end{eqnarray}
The vector $\vec{\alpha} = \{\alpha_x,\alpha_y,\alpha_z\}$ characterizes all non-local properties of $\etil as$. It is sufficient to restrict the $\alpha$'s to $0\le\alpha_{x,y,z}\le\pi/4$, and the set of distinct $\vec{\alpha}$, up to symmetries, make up the so-called Weyl chamber. The complete characterisation of two-qubit interactions that allow for universal ADQC is summarised in the following theorem:

\begin{theorem} \label{t-charach} The global interactions, $\etil a s$, between any two system and ancilla qubits which enable stepwise deterministic and unitary evolution that also admits standardisation procedure for universal ADQC are locally equivalent to (i.e. $D_{as}$ is of the form) 
\AR{\ctR Z ~~{\rm and}~~ \ctRZSWAP.}
\end{theorem}
 
Note that, both $\ctR Z$ and $\ctRZSWAP$ are among the so called ``maximally entangling'' operators, hence the importance of the above theorem is in the \emph{only if} part which proves for the first time the necessity of this particular type of building blocks. 

The proof of Theorem 3 is lengthy and we have broken the characterisation of the non-local operation $D_{as}$ into several lemmas on unitarity, determinism, standardisation and universality. 

\subsection{Conditions for unitarity}

\begin{figure}
   \begin{center}
	\includegraphics[width=0.4\textwidth]{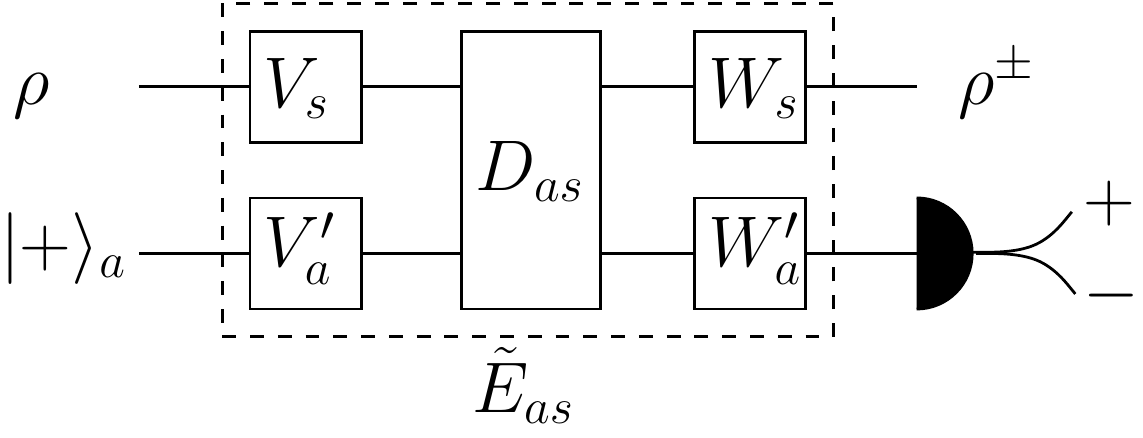}
	\caption{\label{fig:AncillaSystem} 
	The qubit $s$ belonging to the system in state $\rho$ and the ancilla qubit, $a$, are coupled via the two-qubit unitary $\etil a s$. For suitably chosen local unitaries $V'_a$ and $W'_a$, the initial ancilla state $V'_a  \, \ket{+}_a =  \ket{+_{\gamma,\delta}}_a$ and the measurement basis $\{ W'^{\dag}_a \, \ket{+}_a, W'^{\dag}_a \, \ket{-}_a\} =\{ \ket{+_{\ta,\phi}}_a, \ket{-_{\ta,\phi}}_a\}$ are such, that the final state of the system $\rho^{\pm}$ is related to the initial state by a single qubit unitary on $s$, $\rho^{\pm} = U^{\pm}_s \, \rho \, U^{\pm \, \dagger}_s$, conditional on the measurement outcome of the ancilla. }
   \end{center}
\end{figure}

In order to guarantee a stepwise deterministic computation we require that the application of a two-qubit interaction $\etil a s$, followed by measurement of the ancilla shall result in an effective unitary evolution of a single system qubit, this is called \emph{unitary condition}. In what follows, we absorb the local unitaries $W'_a$ and $V'_a$ in Equation (\ref{eq:decomp}) into the preparation and measurement of the ancilla, by choosing appropriate initial state and the measurement basis for ancilla (see Figure~\ref{fig:AncillaSystem}):
\begin{eqnarray*}
V'_a (\ket{+}_a) &=&  \ket{+_{\gamma,\delta}}_a \\
\{ W'^{\dag}_a (\ket{+}_a) \,,\, W'^{\dag}_a (\ket{-}_a) \} &=& \{ \ket{+_{\ta,\phi}}_a \,,\, \ket{-_{\ta,\phi}}_a \}
\end{eqnarray*}

\begin{lemma} \label{lemma:unitarity}
The interaction $\etil a s$ satisfies the unitary condition only if at least one $\alpha_i$ in Equation (\ref{e-Dformula}) is zero. We chose without loss of generality $\alpha_z=0$. For $\alpha_x, \alpha_y \not = 0$ it is necessary and sufficient that both, the initial state and the measurement basis of the ancilla lie in the $X-Y$ plane, i.e. $\gamma=\theta=\pi/2$. When one $\alpha$ is non-zero, w.l.o.g. $\alpha_x \not =0$ and $\alpha_y = 0$, then the ancilla parameters must obey the following relation:
\begin{equation*}
\sin  \theta  \cos \gamma  \sin \phi = \cos \theta  \sin \gamma \sin \delta \, .
\end{equation*}
\end{lemma}

\noindent\textbf{Proof.} Depending on the measurement outcome, denoted by $+$ or $-$, the system qubit is transformed according to
\begin{equation} \label{eq:cp-map}
	\rho \mapsto \rho^{\pm} = { \tilde{\KO}^{\pm}_s \, \rho \, \tilde{\KO}^{\pm \, \dagger}_s  
							\over \tr[ \tilde{\KO}^{\pm}_s \, \rho \, \tilde{\KO}^{\pm \, \dagger}_s ]},
\end{equation}
where the Kraus operators $\tilde{\KO}^{\pm}_s$ are given by
\begin{equation}
	\tilde{\KO}^{\pm}_s = {}_a\bra{\pm}\, \etil a s \, \ket{+}_a
			 = _a\bra{\pm_{\ta,\phi}} \,  W_s  \, D_{as} \, V_s \, \ket{+_{\gamma,\delta}}_a.
\end{equation}
These Kraus operators shall be proportional to unitaries.

The local unitaries on the system qubit, $W_s$ and $V_s$, do not affect whether or not the interaction $\etil a s$ generates a unitary transformation on the system qubit. We will thus focus on the non-local portion $D_{as}$ of $\etil a s$ for now and reintroduce the local unitaries when discussing the standardisation procedure in Lemmas \ref{lemma:standardH} and  \ref{lemma:standardI}. The Kraus operator corresponding to $D_{as}$ alone are 
\begin{equation}
	\KO^{\pm}_s = _a\bra{\pm_{\ta,\phi}} \, D_{as} \, \ket{+_{\gamma,\delta}}_a
\end{equation}
and $D_{as}$ can be diagonalized in the Bell basis \cite{ZVSW2003},
\begin{eqnarray*}
	\ket{\Phi_1} &= \frac{1}{\sqrt{2}} (\ket{00} + \ket{11}), &
	\ket{\Phi_2}   = \frac{-i}{\sqrt{2}} (\ket{00} - \ket{11}) \\
  	\ket{\Phi_3} &= \frac{1}{\sqrt{2}} (\ket{01} + \ket{10}), &
	\ket{\Phi_4}   = \frac{-i}{\sqrt{2}} (\ket{01} - \ket{10}),
\end{eqnarray*}
as 
\begin{eqnarray}
	D_{as}	&=& \sum_{j=1}^4 \, e^{-i \eta_j} \, | \Phi_j \> _{as}\< \Phi_j|, 
\end{eqnarray}
with 
\begin{eqnarray*}
	\eta_1	&= +\alpha_x-\alpha_y+\alpha_z, & 
	\eta_2         = -\alpha_x+\alpha_y+\alpha_z \\
  	\eta_3      &= +\alpha_x+\alpha_y-\alpha_z, &
	\eta_4        = -\alpha_x-\alpha_y-\alpha_z.
\end{eqnarray*}

The completely positive, trace preserving map on $\rho$ in Equation~(\ref{eq:cp-map}) shall have only two possible outcomes, i.e. 
\begin{equation} \label{eq:inanc}
	\mathbbm{1}_s = \KO^+_s \KO^{+\dagger}_s + \KO^-_s \KO^{-\dagger}_s.
\end{equation}
The right hand side of the above equation for a general $D_{as}$ and initial ancilla state $ \ket{+_{\gamma,\delta}}_a$ is equal to
\begin{equation} 
	\left[
	\begin{array}{cc}
  		1+ t & r\\
		r^*  & 1 - t
	\end{array}
	\right],
\end{equation}
where 
\begin{eqnarray}
t &=& \sin 2\alpha_{x} \, \sin 2 \alpha_{y} \, \cos \gamma\\
r &= & \sin \gamma \, \sin 2 \alpha_{z} \, (\sin 2 \alpha_{y} \, \cos \delta - i \sin 2 \alpha_x \sin \delta).
\end{eqnarray}
Since $t=r=0$ must hold, at least one of the $\alpha_j$'s must vanish.

Let us without loss of generality choose $\alpha_z = 0$ and hence $r=0$. If only one $\alpha$ is non-zero then $t=0$ is already true. However, if both $\alpha_x, \alpha_y \not = 0$, then the initial ancilla state must lie in the $X-Y$ plane, i.e. $\gamma = {\pi \over 2}$. These two types of interaction are much studied in physics due to their significance for coupled spin systems. The case with only one non-zero $\alpha$ is referred to as an \emph{Ising interaction}, and the case with two non-zero $\alpha$'s as a \emph{Heisenberg interaction} \cite{spinchains}. Both cases must have individual Kraus operators that are themselves proportional to unitaries,
\begin{equation} \label{eq:partialunitary}
	\KO^{\pm \, \dag}_s \KO^{\pm}_s = p_{\pm} \, \mathbbm{1}_s,
\end{equation}
with probabilities $p_+ + p_- =1$. This puts requirements on the initial state and measurement basis of the ancilla as we discussed next.

For the Heisenberg interaction with $\alpha_z = 0, \alpha_x \not = 0, \alpha_y \not = 0$, the initial state of the ancilla is in the $X-Y$ plane ($\gamma=\pi/2$) and Equation~(\ref{eq:partialunitary}) makes it necessary that the measurement basis of the ancilla must lie in the same plane, i.e. $\theta = \pi/2$. The Kraus operators can then be written as,  up to insignificant phase factors,
\begin{eqnarray} \label{eq:KOHeisenberg}
	\KO^{\pm}_s &=& 
		\left[
		\begin{array}{cc}
			a_{\pm} 		& - b_{\pm} \\
			- b_{\pm}^* 	& - a_{\pm}^*
		\end{array}
		\right]
\end{eqnarray}
with 
\begin{eqnarray} \label{eq:bs&ds}
	\begin{split}
	a_{-}&= \sin \alpha_x \sin \alpha_y \cos {\delta - \phi \over 2} \, - i \cos \alpha_x \cos \alpha_y \sin {\delta - \phi \over 2}, \\
	b_{-}&= \sin \alpha_x \cos \alpha_y \sin {\delta + \phi \over 2} \, + i \cos \alpha_x \sin \alpha_y \cos {\delta + \phi \over 2}, \\
	a_{+}&= \sin \alpha_x \sin \alpha_y \sin {\delta - \phi \over 2} \, + i \cos \alpha_x \cos \alpha_y \cos {\delta - \phi \over 2}, \\
	b_{+}&= \sin \alpha_x \cos \alpha_y \cos {\delta + \phi \over 2} \, - i \cos \alpha_x \sin \alpha_y \sin {\delta + \phi \over 2}.
	\end{split}
\end{eqnarray}
The probabilities for obtaining the measurement results $+$ and $-$  are independent of the system $\rho$ and are given by 
\begin{eqnarray*}
	p_{\pm} &=& {1\over 2} \left(1 \pm  \cos 2 \alpha_x \sin \delta \sin \phi \pm \cos 2 \alpha_y  \cos \delta \cos \phi \right).
\end{eqnarray*}

For the Ising interaction with $\alpha_z = \alpha_y = 0, \alpha_x \not = 0$ since only one of the $\alpha_j$'s is non-zero, then \emph{any initial state of the ancilla}, $ \ket{+_{\gamma,\delta}}_a$, will lead to unitary Kraus operators for the system qubit provided the measurement basis for the ancilla, $\{\ket{+_{\ta,\phi} }_a, \ket{-_{\ta,\phi} }_a\}$, is chosen such that  
\begin{equation} \label{eq:unitary}
	\sin  \theta  \cos \gamma  \sin \phi = \cos \theta  \sin \gamma \sin \delta.
\end{equation}
This can be satisfied, for instance, by $\theta =\gamma$ and $\phi =\delta$ for arbitrary initial state $\ket{+_{\gamma,\delta}}_a$. The Kraus operators for the Ising interaction are given by
\begin{eqnarray} \label{eq:KOIsing}
	\KO_s^{\pm} &=& A_{\pm} \,  \mathbbm{1} + i \, (-1)^{n_{\pm}} \, B_{\pm} \, X
\end{eqnarray}
where $n_{\pm}$ are arbitrary integer numbers and the real coefficients $A_\pm, B_\pm$  are
\begin{eqnarray*}
	A_{\pm} &=& {\cos \alpha_x \over \sqrt 2} \,\sqrt{1 \pm \cos \gamma \cos \theta \pm \sin \gamma \sin \theta \cos (\delta-\phi) }, \\
	B_{\pm} &=&  {\sin \alpha_x \over \sqrt 2}
 \sqrt{1\mp \cos \gamma \cos \theta  \pm \sin \gamma \sin \theta \cos (\delta+\phi)}, 
\end{eqnarray*}
where the angle $\theta$ and the phase $\phi$ of the measurement basis has to satisfy Equation~(\ref{eq:unitary}). 
The probabilities for the two measurement outcomes are
\begin{eqnarray*}
	p_{\pm} &=& {1 \over 2} \left(1 \pm \sin \theta  \sin \gamma \cos \delta \cos \phi 
	\pm  \cos 2 \alpha_x (\cos \theta \cos \gamma + \sin \theta \sin  \gamma  \sin \delta \sin \phi) \right).
\end{eqnarray*}
This concludes the proof of lemma. \qed

\subsection{Conditions for correctable branching}
  
As the measurement results are random, the unitary operation that is applied to the state is either  $U^{+}_s $ or $U^{-}_s$ where $U^{\pm}_s = {\KO^{\pm}_s / \sqrt{p_{\pm} } }$ are the normalised Kraus operators. To generate a deterministic evolution, for example with $U^+_s$, we require that the other unitary, $U^-_s$, can be corrected with an additional Pauli correction $P_s$ (up to an unimportant relative phase $\Delta$), 
\begin{equation}
	U^-_s =  e^{i \Delta}  \, P_s \, U^+_s.
\end{equation}
General Pauli corrections are of the form $P_s(a,b,c) = a X_s + b Y_s + c Z_s$ with $a, b, c \in \mathbbm{R}$ and $a^2 +b^2+c^2=1$ and have the properties
\begin{equation}
P = P^{\dag}, ~~ P^2 = \mathbbm{1}, ~~\tr[P] =0.\nonumber
\end{equation}
Pauli corrections are the standard choice in measurement-based computation. 
Their structure is well understood and essential in ensuring an overall deterministic computation. 
Reuniting the branches with a Pauli correction after the measurement places no direct constraint on the interaction,  $D_{as}$. Instead this can be achieved by choosing the initial state and the measurement basis of the ancilla appropriately, as specified below.

In the Heisenberg case the correctable Kraus operators \eqref{eq:KOHeisenberg} must fulfil the relation $a_{+}^* a_{-} + b_{+} b_{-}^* =0$, i.e. the two non-zero $\alpha$'s must relate to each other as 
\begin{equation}
	\cos 2 \alpha_x \tan \delta = \tan \phi \cos 2 \alpha_y.
\end{equation}
It can be seen that all values of $\alpha_x$ and $\alpha_y$ can be covered independently by choosing the phases $\delta$ and $\phi$ for the ancilla appropriately. 

However, if one of the interaction parameters is chosen as $\alpha_y=\pi/4$ the above relation implies $\delta=0$ and the probability for each branch becomes $p_{\pm} = {1 \over 2}$. It will be discussed in Lemma \ref{lemma:standardH} that this requirement is indeed necessary to allow for the standardisation of a computational pattern. The Kraus operators Equation~(\ref{eq:KOHeisenberg}) then have the coefficients 
\begin{eqnarray} \label{eq:KH1}
	\begin{split}
	a_{-}&= {1 \over \sqrt{2}} 
		\left(\sin \alpha_x \cos {\phi \over 2} + i \cos \alpha_x \sin {\phi \over 2}\right) = b^*_+, \\
	b_{-}&= {1 \over \sqrt{2}} 
		\left(\sin \alpha_x \sin {\phi \over 2} + i \cos \alpha_x \cos {\phi \over 2}\right) = - a^*_+.
	\end{split}
\end{eqnarray}
It can be seen that these parameters imply that the Pauli correction between the two Kraus operators is $Y$ (up to a phase), $\KO^+_s = i \, Y_s \, \KO^-_s$. 

In the Ising case, the Kraus operators are a sum of identity and Pauli-X only, see Equation~(\ref{eq:KOIsing}) and the only non-trivial correction available is $X$. To fulfil ${\KO^+_s \over \sqrt{p_+}} = X_s  {\KO^-_s \over \sqrt{p_-}}$, one has to choose the ancilla's initial state and measurement basis such that
\begin{equation} \label{eq:IPcorr}
	A_+ A_-  +(-1)^{n_+ + n_-}B_+ B_- =0.
\end{equation}
It can be seen that one $n$ has to be even and the other one odd in oder to satisfy this condition. The probabilities for the two measurement outcomes need not be balanced and are given as $p_{\pm}= A_{\pm}^2 +B_{\pm}^2$. The interaction strength $\alpha_x$ can then be manipulated by choosing the ancilla parameters,
\begin{equation}
	\tan^2 \alpha_x  =\sqrt{1-\left(\cos \gamma \cos \theta + \sin \gamma \sin \theta \cos (\delta-\phi) \right)^2 \over 
	1-\left(\cos \gamma \cos \theta - \sin \gamma \sin \theta \cos (\delta+\phi) \right)^2}.
\end{equation}

\subsection{Conditions for standardisation} 
  
A computation can be standardised, see \ref{sec:adqc-standard}, when a general Pauli correction transforms, under the interaction $D_{as}$, into a tensor product between ancilla and system qubit, where the new correction on the system qubit is again a general Pauli operation. By symmetry of $D_{as}$ the same must be true for the ancilla, i.e. we require
\begin{equation} \label{eq:standard}
	D_{as} \,\,\,\, \mathbbm{1}_a \otimes P_s (a,b,c) \,\,\,\, D^{\dag}_{as} = T_a \otimes Q_s,
\end{equation}
where $T_a$ and $Q_s$ are either general Pauli operations or the identity operation. If this relation is valid the correction $P_s$ can be commuted through future interactions and shifted to the very end of the computation.

\begin{lemma} \label{lemma:standardH}
For the Heisenberg interaction there are two possible solutions for pairs of unitaries $D_{as}(\alpha_x,\alpha_y,0)$ and Pauli corrections $P(a,b,c)$, up to relabeling of the $\alpha$'s, that obey the standardisation relation Equation~(\ref{eq:standard})\\[2ex]
       1. Fixed Heisenberg interaction: $\alpha_y=\pi/4$ and $\alpha_x=\pi/4$.      \\
       \indent Here Pauli corrections $P(a,b,0)$ and $P(0,0,1)$ transform according to\\
       \indent $\begin{array}{lcl}
       \mathbbm{1}_a \otimes (a X_s + b Y_s)
               & \mapsto & (a Y_a - b X_a) \otimes Z_s, \quad a^2 +b^2 =1,\\
       \mathbbm{1}_a \otimes Z_s
               & \mapsto & \hspace{10ex} Z_a \otimes \mathbbm{1}_s.
       \end{array}$  \\[2ex]
       2. General Heisenberg interaction: $\alpha_y=\pi/4$ and $\alpha_x\not=\pi/4$. \\
       \indent Pauli corrections $P(1,0,0)$ transform according to\\
       \indent $\begin{array}{lcl}
       \mathbbm{1}_a \otimes X_s
               & \mapsto & Y_a \otimes Z_s.
       \end{array}$
\end{lemma}

\noindent\textbf{Proof.} For $D_{as} ( \alpha_x, \alpha_y,0)$ the general Pauli correction $P(a,b,c)$ transforms according to
\begin{eqnarray} \label{eq:DH}
\begin{split}
D_{as}   \mathbbm{1}_a \otimes P_s    D^{\dag}_{as}
               &=&  \mathbbm{1}_a \otimes (a \cos 2 \alpha_y X_s + b \cos 2 \alpha_x Y_s + c \cos 2 \alpha_x \cos 2 \alpha_y Z_s) \\
               & & + \sin 2 \alpha_x \, X_a \otimes (-b Z_s+ c \cos 2 \alpha_y Y_s) \\
               & & + \sin 2 \alpha_y \, Y_a \otimes (a Z_s- c \cos 2 \alpha_x X_s) \\
               & & + \sin 2 \alpha_x \sin 2 \alpha_y \, Z_a \otimes c \mathbbm{1}_s.
\end{split}
\end{eqnarray}
To determine which interactions fulfil Equation~(\ref{eq:standard}) we take the partial traces of either the ancilla or the system. Reading from the right hand side of Equation~(\ref{eq:standard}) the traces must be, up to global phases,
\begin{eqnarray}  \label{eq:trick}
 \tr[T_a]  \cdot Q_s = \left\{
                       \begin{array}{cl}
                               2 Q_s  & \mbox{iff } T_a = \mathbbm{1}\\
                               0  & \mbox{iff } T_a \mbox{ is a general Pauli operation},\\
                       \end{array} \right. \\
 T_a \cdot \tr[Q_s] = \left\{
                       \begin{array}{cl}
                               2 T_a  & \mbox{iff } Q_s = \mathbbm{1}\\
                               0  & \mbox{iff } Q_s \mbox{ is a general Pauli operation}.\\
                       \end{array} \right.
\end{eqnarray}
The partial trace over the ancilla in the Heisenberg case, Equation~(\ref{eq:DH}), is
\begin{equation}
       \tr_a[D_{as}     \mathbbm{1}_a \otimes P_s    D^{\dag}_{as}]
       = 2 \cdot (a \cos 2 \alpha_y X_s + b \cos 2 \alpha_x Y_s + c \cos 2 \alpha_x \cos 2 \alpha_y Z_s).
\end{equation}
For $\alpha_x, \alpha_y \not = 0$ this expression can never be equal to $2 Q_s$ where $Q_s$ is a general Pauli or the identity for the system qubit. However, the right hand side can vanish (and $T_a$ must hence be a general Pauli) when $a=0$ and $\alpha_x=\pi/4$ or $b=0$ and $\alpha_y=\pi/4$ or  $\alpha_x=\alpha_y = \pi/4$. Thus, for the Heisenberg interaction one of the parameters has to be fixed to $\pi/4$ to enable full standardisation of the computation and we choose, without loss of generality, $\alpha_y=\pi/4$.
In the case that $b=0$, i.e. for corrections $P(a,0,c)$, the right hand side of Equation~(\ref{eq:DH}) becomes
\begin{equation}
       Y_a \otimes (a Z_s - c \cos \alpha_x X_s) + c \sin 2 \alpha_x \, Z_a \otimes \mathbbm{1},
\end{equation}
which implies either $c =0$ to match condition Equation~(\ref{eq:trick}). When both $\alpha=\pi/4$ then the right hand side of Equation~(\ref{eq:DH})  becomes
\begin{equation}
       (-b X_a + a Y_a) \otimes Z_s + c Z_a \otimes \mathbbm{1},
\end{equation}
and the correction must thus be either $P(a,b,0)$ or $P(0,0,1)$. \qed

An example for Case 1 is the $\ctRZSWAP$ interaction whereas Case 2 allows a range of parameters $0 < \alpha_x < \pi/4$; however, only one type of correction, $X$, obeys the standardisation condition. This becomes an issue when composing several ancilla-driven operations after one another. 

\begin{figure}
   	\begin{center}
	\includegraphics[width=0.8\textwidth]{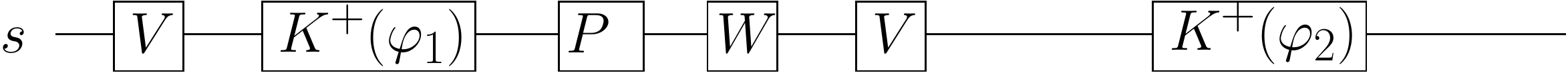}\\[2ex]
	\includegraphics[width=0.8\textwidth]{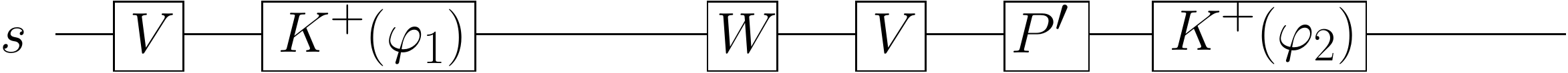}\\[2ex]
	\includegraphics[width=0.8\textwidth]{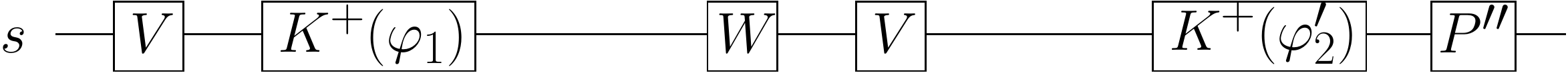}
	\caption{\label{fig:composing} 
	Composition of gates on a single qubit $s$. In this example, each line represents the same effective computation. Attempting to implement $\KO^+ (\varphi_1)$ results, with probability $p_-$, in the actual operation $\KO^- (\varphi_1) = P \, \KO^+ (\varphi_1)$  where a  correction $P$ appears (first line). The correction can be shifted past local unitaries $W$ and $V$ (second line) and past subsequent ancilla-driven operations $\KO^+ (\varphi_2)$ (third line), see the text for details. }
	\end{center}
\end{figure}

\begin{lemma}
Case 2 in Lemma \ref{lemma:standardH} is not compositional. 
\end{lemma}
\noindent\textbf{Proof.} In order to implement arbitrarily complex computations it is necessary to repeatedly apply ancilla-driven unitary transformations to a single system qubit, see Figure~\ref{fig:composing}. 

The full ancilla-system interaction $\etil a s$ plus measurement of the ancilla generates the effective operator sequence $\bra{\pm}_a \, \etil a s  \, \ket{+}_a =  W_s \, \KO^\pm_s (\varphi)\, V_s$ on the system qubit, where $W_s$ and $V_s$ see Equation~(\ref{eq:decomp}) are the local unitaries on the system qubit and $\varphi$ denotes all ancilla parameters. Due to the probabilistic nature of the measurement, when attempting to implement the operation $\KO^+ (\varphi_1)$, the actual outcome could be $\KO^- (\varphi_1) \propto P \,\KO^+ (\varphi_1)$ with a correction $P$, pictured in the first line, Figure~\ref{fig:composing}. The branching behaviour discussed in the previous subsection requires that $P = Y$ for the Heisenberg interaction and $P = X$ for the Ising interaction. The local unitaries affect this original correction and transform it into a possibly different general Pauli $P'$, as shown in the second line,  Figure~\ref{fig:composing}. 

Constructing an arbitrary computation requires composition of subsequent ancilla-driven operations, $\KO^+ (\varphi_2)$ etc. When commuting the correction $P'$ with the $\KO^+ (\varphi_2)$ the result must be a new Kraus operation $\KO^{\pm} (\varphi_2')$, implemented by a (possibly) different set of ancilla parameters $\varphi_2'$, and a new general Pauli correction $P''$ after the new Kraus operation, as depicted in the third line of Figure~\ref{fig:composing}. In the two cases of interest here, cases 2 and 4, Pauli $X$ is the only allowed correction, i.e. $P' =X$. Moreover, in order for the new correction $P''$ to be shifted further, it has to be identical to the branching relation $P$, or $\mathbbm{1}$. The pictorial shifting conditions imply the following mathematical relations
\begin{eqnarray} \label{eq:shifting}
	\begin{split}
	V \, W  \,	& P 	= \, e^{i \Delta} &P' \, \, \, \, \, \, \, \, \, \, \, \, V \, W, \\
	{ \KO^+(\varphi_2) \over \sqrt{p_+(\varphi_2)}} \, & P' 	= \, e^{i \Delta'} & P'' \, { \KO^{\pm}(\varphi'_2) \over \sqrt{p_{\pm}(\varphi'_2)}}, \\
	\, & P'' = \, e^{i \Delta''} &P \mbox{ or } e^{i \Delta''} \, \, \mathbbm{1},
	\end{split}
\end{eqnarray}
where $\Delta, \Delta', \Delta''$ are arbitrary global phases.

In Case 2 of Lemma \ref{lemma:standardH} this set of equations \emph{cannot} be fulfilled, as there is no pair of ancilla parameters $\varphi_2$ and $\varphi'_2$ for which the Kraus operators Equation~(\ref{eq:KOHeisenberg}) with coefficients Equation~(\ref{eq:KH1}) obey $K^+(\varphi_2) \, Y = e^{i \Delta''} X \, K^{\pm}(\varphi_2')$ or $K^+(\varphi_2) \, Y = e^{i \Delta''} \, K^{\pm}(\varphi_2')$. This implies that the possible ancilla-driven computations using a general Heisenberg interaction are extremely limited - only one ancilla interaction per system qubit can be tolerated.  To enable composition with the Heisenberg interaction forces the interaction parameter to take the special value $\alpha_x=\pi/4$. \qed

Similarly, we characterise the Ising interaction cases.

\begin{lemma} \label{lemma:standardI}
For the Ising interaction there are two possible solutions for pairs of unitaries $D_{as}(\alpha_x,0,0)$ and Pauli corrections $P(a,b,c)$, up to relabeling of the $\alpha$'s, that obey the standardisation relation Equation~(\ref{eq:standard})\\[2ex] 
	3. Fixed Ising interaction: $\alpha_x=\pi/4$. 	\\
	\indent Here Pauli corrections $P(1,0,0)$ and $P(0,b,c)$ transform according to\\
	\indent $\begin{array}{lcl}
	\mathbbm{1}_a \otimes X_s & \mapsto & \mathbbm{1}_a \otimes X_s, \\
	\mathbbm{1}_a \otimes (b Y_s +c Z_s) & \mapsto & X_a \otimes (b Z_s - c Y_s),  \quad a^2 +b^2 =1.
	 \end{array}$ \\[2ex]
	4. General Ising interaction: $\alpha_x\not=\pi/4$. \\
	\indent Pauli corrections $P(1,0,0)$  transform according to\\
	\indent $\begin{array}{lcl}
	\mathbbm{1}_a \otimes X_s & \mapsto & \mathbbm{1}_a \otimes X_s. \\
	 \end{array}$ 
\end{lemma}

\noindent\textbf{Proof.} The argument is similar to the proof of the previous lemma. \qed

Case 4 again allows only one type of correction, an $X$ correction. We prove next that this requirement restricts any obtained ancilla-driven unitaries on a single system qubit to be on a fixed plane and thus fails universality.

\subsection{Universality}

\begin{figure}
   	\begin{center}
	\includegraphics[width=0.95\textwidth]{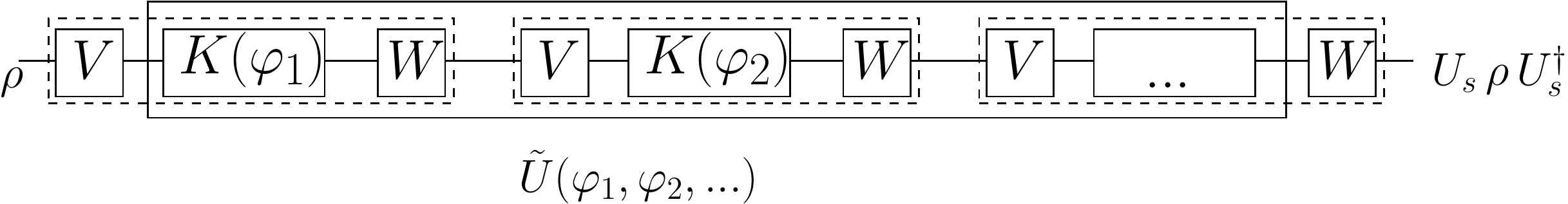}
	\caption{\label{fig:universalI}
	Any sequence, $U(\varphi_1, \varphi_2, \dots) = W \, \tilde{U}(\varphi_1, \varphi_2, \dots) \, V$,  of $\KO$ operations and $VW$ operations commutes or anti-commutes with $X$.}
	\end{center}
\end{figure}

The general Ising interaction, Case 4, fulfils the set of equations \ref{eq:shifting} as the Pauli $X$ correction commutes with the actual Kraus operators
\begin{equation*}
\KO^+(\varphi_2) \, X = X \, \KO^{+}(\varphi_2)
\end{equation*} 
for $\KO^{+} = A_{+} \,  \mathbbm{1} + i \, (-1)^{n_{+}} \, B_{+} \, X$, given in Equation~(\ref{eq:KOIsing}), where the product of the local unitaries has to obey $V W  X  = e^{i \Delta} X V  W$, i.e. $V W$ is either of the form (up to global phases) $V W= a \mathbbm{1} + i b X$ or of the form $V W= a Y + b Z$ with some coefficients $a,b \in \mathbbm{R}$ and $a^2 +b^2=1$. However, this leads to the following lemma.

\begin{lemma}
Case 4 in Lemma \ref{lemma:standardI} is not universal. 
\end{lemma}
\noindent\textbf{Proof.} Any sequence of consecutive application of $\KO (\varphi)$ and $VW$ (see Figure~\ref{fig:universalI})
\begin{eqnarray*}
U (\varphi_1, \varphi_2, \dots) =  \dots W \, \KO(\varphi_3) \, V \quad W \, \KO(\varphi_2) \, V \quad  W \, \KO(\varphi_1) \, V
\end{eqnarray*}
contains a kernel, $U  (\varphi_1, \varphi_2, \dots) = W \, \tilde{U} (\varphi_1, \varphi_2, \dots) \, V$, that is again of the form 
\begin{eqnarray*}
	\phantom{\mbox{or } \, } \tilde{U} (\varphi_1, \varphi_2, \dots) = a(\varphi_1, \varphi_2, \dots) \, \mathbbm{1} + i b (\varphi_1, \varphi_2, \dots) \, X \\
	\mbox{or } \,	\tilde{U} (\varphi_1, \varphi_2, \dots) = a(\varphi_1, \varphi_2, \dots) \, Y + b (\varphi_1, \varphi_2, \dots) \, Z.
\end{eqnarray*}
The coefficient functions $a(\varphi_1, \varphi_2, \dots)$ and $b(\varphi_1, \varphi_2, \dots)$ with $a^2+b^2=1$ will depend on the choice of all ancilla parameters $\varphi_1, \varphi_2, \dots$ in the sequence. It is easy to show that both classes of operations, for arbitrary parameters, can move a given input state only in one plane of the Bloch sphere,  parallel to the $Y-Z$ plane. Finally, we note that the additional first and last local unitary operation, $V$ and $W$, which individually can take  arbitrary form as long as their product $VW$ \mbox{(anti-)commutes} with $X$, are \emph{fixed}. That means that once $W$ an $V$ are given, the plane where $\tilde{U}$ operates is tilted into an arbitrary, yet fixed direction in which the state can evolve. Since it is impossible to move out of that plane it is implied that the $U (\varphi_1, \varphi_2, \dots)$ for general Ising interaction, Case 4, does not allow universal single qubit preparation. \qed

The final step for the proof of Theorem \ref{t-charach} is given below. 

\begin{lemma}
The remaining cases 1 and 3 are universal. 
\end{lemma}

\noindent\textbf{Proof.} It suffices to identify suitable local operations that ensure universality. The $\ctR Z$ interaction is an example for Case 3 and choosing the local unitaries $V_s$ and $W_s$ appropriately, for instance: 
\begin{eqnarray*}
\etil as &=& H_a \otimes H_s \, \ctR Z_{as} \\
         &=& H_a \, P(\pi/2)_a \, H_a \otimes H_s \, P(\pi/2)_s \, H_s \,\,\, D_{as}(\pi/4,0,0) \,\,\, H_a \otimes H_s 
\end{eqnarray*}
makes this model universal as we proved in Subsection \ref{sec:universal}. The $\ctRZSWAP$ interaction is an example for Case 1 and it is easy to verify that ADQC model with $\ctRZSWAP$ is indeed the same as the one-way model, as the position of ancilla and system qubits are swapped, and hence universal. \qed

Note that the local unitaries on the ancilla are irrelevant as they just redefine initial state and measurement basis of the ancilla. Moreover, as said before, for the Case 3 the local unitaries on the system have to enable a ``change of basis'', this would rule out the class of diagonal unitaries. However, the full characterisation of local unitaries that makes each cases 1 and 3 universal remains an interesting open problem. For the rest of this paper we will focus on the $\etil as = H_a \otimes H_s \, \ctR Z_{as}$ interaction which defines a new class of entangled states.

\section{Twisted Graph States}\label{s-twist}

The main issue with the $\etil i j$ operators is the fact that they are
non-commuting, therefore after standardisation their order will be
important. Another important property of an ancilla-driven pattern is
that system qubits interact only with ancilla qubits. Therefore we
introduce a multipartite entangled state as a graph over ancilla and
system qubits and $\etil i j$ edges with extra condition to address
the above mentioned requirements. In Section \ref{sec:trans-mbqc} we
show how this new class of states can be viewed as open graph states up
to some local swap operators. This is the reason behind the chosen
name.

\begin{figure}[h]
\begin{center}
\includegraphics[scale=0.35]{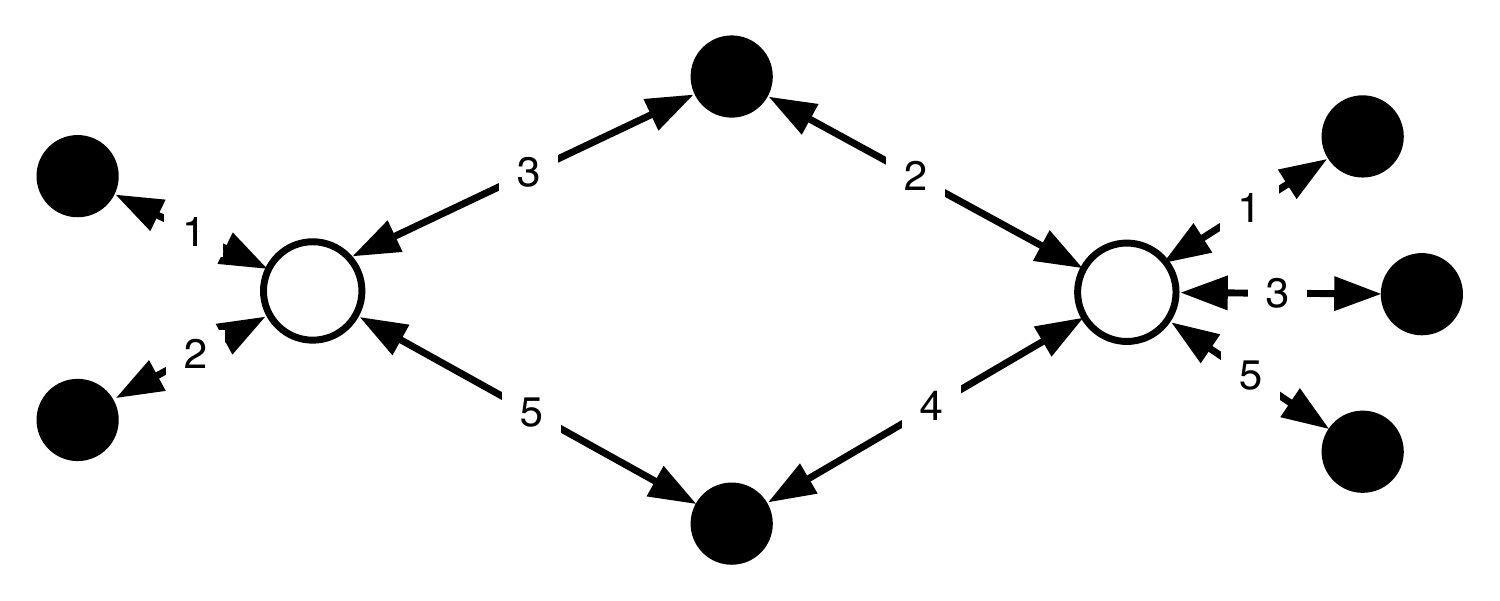}
\caption{An open twisted graph state where the system qubits are the
  white circles that will not be measured and the rest are ancilla
  qubits. The edges are $\etil {}{}$ interactions and edge labels
  denote the partial order.}
\label{f-tildgraph}
\end{center}
\end{figure}

\begin{definition} An \emph{open twisted graph state} $(G,S,A,\mathcal{C})$ consists
of a bipartite graph $G$ over disjoint sets of qubits $S$ and $A$,
called \emph{systems} and \emph{ancillas}, such that the maximum degree of any
ancilla node is 2, together with an edge labelling $\mathcal C$. The labels define a partial ordering over 
edges where the order is total and strict on any edges that share a common vertex, \ie~it defines an edge colouring. 

The corresponding quantum state, denoted as
$\ket{\widetilde{E}_G}$, is obtained by preparing qubits in $S$ in given
arbitrary states and all the qubits in $A$ in the $\ket +$ state, and then
applying $\etil i j$ over corresponding qubits according to the partial
order of $\mathcal C$ (see Figure \ref{f-tildgraph}).  \end{definition}

One may think of an open twisted graph state as the beginning of the
definition of an ancilla-driven pattern, where one has already decided
how many qubits will be used ($V=S\cup A$) and how they will be
entangled: 
\AR{ \widetilde{E}_G:=\prod_{\{i,j\}\in \tilde {E}} \etil  ij .} 
To complete the definition of the pattern it remains to
decide which angles will be used to measure ancilla qubits and, most
importantly if one is interested in determinism, which dependent
corrections will be applied. Conversely, any ancilla-driven pattern
has a unique underlying open twisted graph state obtained by
forgetting measurements and corrections where the colour of the edges
is given by the most partial order of the $\etil i j$ commands which
respects the non-commuting order. Recall that two $\etil i j$ commands commute if and only if they act on disjoint set of qubits. Note that
the depth of this partial order is the true depth of the preparation of the state.

Different ordered colourings over the same graph structure
might lead to different twisted graph states and consequently different
patterns of computation. We leave as an open question whether one can
find a more relaxed definition that can still uniquely define the
entangled state corresponding to an ancilla-driven pattern. On the
other hand our restrained definition will allow us to derive the
dependency structure of the measurements directly from the order of
the colouring, this is the topic of the next subsection.  

\subsection{Dependency structure}
\label{sec:depend}

We will use the graph stabiliser formalism~\cite{NC00,graphstates} to
construct a deterministic pattern. Recall that for any open graph
state $\ket{E_G}$ defined over a graph $G$ with vertices $V$, we have
the following set of equations for all the non-input qubits: \AR{
  X_{i}\prod_{j\in G(i)}Z_j (\ket {E_G}) = \ket {E_G} } where $G(i)$
is the set of neighbour vertices of $i$ in $G$. The above Pauli
operators are called the stabiliser operators of $\ket {\psi}$.

Similarly we define the stabiliser operators of a given twisted graph
state $ \ket {\widetilde{E}_G}$ defined over a graph $G$ with vertices
$V=S\cup A$. We will use the following notation as well. Define
$S(a)$ for $a \in A$ to be the attached system qubit $s \in S$ with
the smallest edge label and $S'(a)$ to be the other one if
it exists; $N(s)$ for $s\in S$ to be the set of ancilla qubits
connected to the system qubit $s$; and finally $G_{S(a)}$ to be the
sub-graph with edges between $S(a)$ and $N_{S(a)}$.

Consider first a simple case where $\widetilde{E}_G = \etil
a{S(a)} N_a^{\ket{+}}$. Then the stabiliser has the form
\EQ{\label{e-stab1} Z_{a}X_{S(a)} (\ket {\widetilde{E}_G}) &=& \ket
  {\widetilde{E}_G} } The above equation is due to \AR{ Z_{a}X_{S(a)}
  (\etil a{S(a)} N_a^{\ket{+}})
  &=& \etil a {S(a)}X_a Z_{S(a)}Z_{S(a)}N_a^{\ket{+}} \\
  &=& \etil a {S(a)} N_a^{\ket{+}} } and for another simple case of
$\ket {\widetilde{E}_G} = \etil a{S'(a)} \etil a{S(a)} N_a^{\ket{+}}$
we have 
\EQ{\label{e-stab2} X_{a}X_{S(a)} (\ket {\widetilde{E}_G}) &=&
  \ket {\widetilde{E}_G} } again due to \AR{ X_{a}X_{S(a)} (\etil
  a{S'(a)}\etil a{S(a)} N_a^{\ket{+}})
  &=& \etil a{S'(a)}\etil a{S(a)}X_a Z_{S(a)}Z_{S(a)}N_a^{\ket{+}} \\
  &=& \etil a{S'(a)}\etil a{S(a)} N_a^{\ket{+}} .}

\begin{figure}[h]
\begin{center}
\includegraphics[scale=0.35]{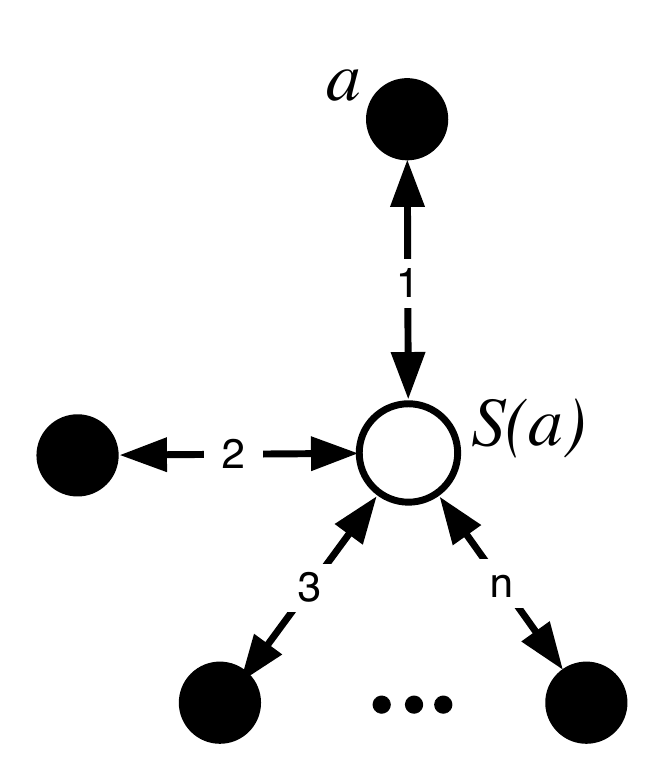}
\caption{The generic case for the local study of stabiliser at ancilla $a$.}
\label{f-depend}
\end{center}
\end{figure}

In order to generalise the above cases the following rewrite rules for
Pauli commutations will be used:
\begin{itemize}
\item Equation \ref{etil-cx}, $\etil i j \cx is =\cx js\cz is\etil
  ij$, transforms the $X$ operation on the system qubit to the next
  immediate ancilla qubit, introducing a $Z$ operation at the system
  qubit.
\item Equation \ref{etil-cz}, $\etil i j \cz is =\cx is\etil ij$,
  replaces the $Z$ operation on the system qubit with an $X$
  operation.
\end{itemize}
Unlike the stabiliser of the graph state which has a local structure,
in the case of a twisted graph state the stabiliser of $a$ may affect the
whole of the graph. This action is, however, recursive and can be
defined using the collection of several local actions. Define the
label of an ancilla node to be the same as the label of the edge $\etil {a}{S(a)}$. Consider an
ancilla qubit $a$ and those qubits in $N(S(a))$ with label greater than label of $a$ (see Figure
\ref{f-depend}). We can assume, without loss of generality, that all
edges connected to $S(a)$ have labels $1$ to $n$ with $1$ being the
label of $a$. This is due to the fact that the stabiliser of $a$ has
an effect only on those qubits in $N(S(a))$ where their edge
interaction are after the edge interaction of $a$ and $S(a)$ hence
having a greater edge label.

\begin{definition} \label{d-localstab}
Given a twisted graph state $(G,S,A,\mathcal{C})$, a \emph{local
  stabiliser} on the ancilla qubit $a$ is defined as following:
\begin{itemize}
\item Consider vertices in $G_{S(a)}$ with labels greater than label of
  $a$ and relabel them from $1$ to $n$ according to $\mathcal{C}$ ordering, with $1$ being the
label of $a$.
\item Add $X$ on vertices in $N(S(a))$ with even label and degree 1.
\item Add $Z$ on vertices in $N(S(a))$ with even label and degree 2.
\item Add $X$ on the system qubit if $n$ is odd otherwise add
  $Z$.
  \end{itemize}   
  We denote the above set of Pauli operators with $P_l(S(a))$ which
  acts on a subset of qubits in $N(S(a))$.  \end{definition}

Define $I(a)$ to be the set of degree-two ancilla qubits in $N(S(a))$
with even label that get a $Z$ Pauli operator in the definition of the
local stabiliser of $a$. The stabiliser of $a$ will have the same
local effect as defined above over $S(a')$ for all the $a' \in I(a)$
and the same structure repeats for vertices in $I(a')$. Therefore we
define recursively such qubits \AR{ I^*(a)=\{a' | \exists n : a' \in
  I^n (a)\} } where $I^1 (a) = I(a)$ and $I^{n+1}(a)=\bigcup_{a'\in
  I^n(a)} I(a')$. We can now present a recursive definition for the
stabiliser of a twisted graph state as the product of a collection of
local stabilisers.

\begin{definition} \label{d-stab} Given a twisted graph state $(G,S,A,\mathcal{C})$,
the \emph{stabiliser} on the ancilla qubit $a$ is defined as
follows: \AR{
  P(a)&=&Z_a\prod_{a'\in I^*(a)} P_l(S(a')) \;\;\;\;
\text{if $a$ has degree one} \\
  P(a)&=&X_a\prod_{a'\in I^*(a)} P_l(S(a')) \;\;\;\; \text{otherwise}
} \end{definition} 
It is a straightforward but cumbersome computation to show the
correctness of the above definition and we omit the details of the proof. 

We can now adapt the notion of flow and generalised flow of graph
states \cite{Flow06,g-flow} for twisted graph states to derive a
sufficient condition for determinism. The key idea is exactly the same
as in the MBQC case based on the following simple observation. We
could make a measurement $\GM {\cplane XY} \alpha i$ ``deterministic''
(corrected) if it could be pre-composed by an anachronical $Z_i^{s_i}$
correction (\ie~conditioned on the outcome of a measurement which
hasn't happened yet). This unphysical scenario is a useful starting
point for our proof.  \AR{ \bra{{+_{\cplane XY, \alpha}}}_a&=&\GM
  {\cplane XY} \alpha a \cz a {m_a} } The flow construction guarantees
that a deterministic pattern with anachronical corrections \AR{
  \mfr {P} &=& \prod^{\mathcal C}_{a\in A} \;\;
\bra{{+_{\cplane XY, \alpha}}}_a\;\; \widetilde{E}_G \\\\
  &=& \prod^{\mathcal C}_{a\in A} \;\; \GM {\cplane XY} \alpha a \cz a
  {m_a} \;\; \widetilde{E}_G } can be transformed into a runnable
pattern, where all dependencies will respect the proper causal
ordering. The key observation which allows us to transform this into a
runnable pattern is that the flow construction defines a stabiliser
$P_{f(a)}$ which when composed with the anachronical correction, forms
an operator which commutes with the measurement, and thus the pattern
can be brought into runnable order.

For simplicity in the rest of the paper we consider only patterns where
degree-two vertices are measured with Pauli $Z$ and degree-one
vertices are measured in the $\cplane XY$ plane, we use the generic term $\GM
{\lambda_a} {\alpha_a} a$ for both cases. This class of patterns are
large enough to introduce a universal ADQC model as they include the
generating pattern introduced in Section \ref{sec:universal}. However
the definition of flow and determinism can be extended to the more
general case as well.

\begin{definition} An open twisted graph state $(G,S,A,\mathcal C)$ has \emph{causal
  flow} if there exists a partial order $>$ over $V$ such that for all $a\in A$ and all
vertices $a' \in P(a)$ we have $a<a'$ except for those $a'$ that will
be measured with Pauli $Z$.  \end{definition}

\begin{theorem} \label{t-deter} Suppose the open twisted graph state
$(G,S,A,\mathcal{C})$ has a causal flow with the partial order
$>$. Define: \AR{
  C(a)&=&P(a)Z_a \;\;\; \text{ for all degree-one ancilla $a$} \\
  C(a)&=&P(a)X_a \;\;\; \text{ for all degree-two ancilla $a$} } then
the pattern: \AR{ \mfr P_{G,\vec{\al}}&:=&\prod^{>}_{a \in A} \;\;
  C(a)^{m_a}\;\;\GM {\lambda_a} {\alpha_a} a \;\; \widetilde{E}_G }
where the product follows the dependency order $>$, is runnable,
uniformly and strongly deterministic.  \end{theorem} 

\noindent\textbf{Proof.} The proof is
based on the following equations first consider the $\cplane XY$ measurement
case for degree-one ancillas: \AR{
\label{e1}
\bra{{+_\al}}_a (\widetilde{E}_G)
&=&\m\al a\cz a{m_a} (\widetilde{E}_G) \\
&=& \m\al a\cz a{m_a} P(a)^{m_a}(\widetilde{E}_G) \\
&=& C(a)^{m_a}\m\al a (\widetilde{E}_G) } Similarly for the $Z$
measurement over degree-two ancillas we have: \AR{ \bra{{0}}_a
  (\widetilde{E}_G)
  &=&\m Z a\cx a{m_a} (\widetilde{E}_G) \\
  &=& \m Z a\cx a{m_a} P(a)^{m_a} (\widetilde{E}_G) \\
  &=& C(a)^{m_a}\m Z a (\widetilde{E}_G) } Hence we can write: \AR{
  \prod^{>}_{a \in A}\bra{{\lambda_a},{\alpha_a}}_a \widetilde{E}_G
  =\prod^{>}_{a \in A} \;\;C(a)^{m_a}\;\;\GM {\lambda_a} {\alpha_a} a
  \;\; \widetilde{E}_G.  } The left hand side is clearly a uniformly
and strongly deterministic pattern. The right hand side pattern is
runnable as the introduced corrections follow the partial order $>$
except for the $Z$ correction introduced over degree-two
ancillas. However one can ignore them since these qubits will be
measured with Pauli $Z$ and we have $M_a^Z \; Z_a^m = M_a^Z$. This
finishes the proof.  \qed

It is interesting to note that the flow definition for a graph state
was based on the geometry of the underlying graph, whereas in a twisted
graph state it is based on the edge colouring order. Indeed, as
mentioned before, different edge colourings lead to different twisted
graph states and hence different flow constructions. Roughly speaking,
the edge colouring plays the role of geometry for the twisted graph
states.

\section{Compositional Embedding}\label{s-embed}

One of the main foci in constructing direct translations between
models is to study parallelism as the more parallel the
computation, the more robust it is against decoherence. Recently the
advantage of MBQC over GBQC in terms of depth complexity has been
demonstrated where a logarithmic separation was shown
\cite{BK06}. We will prove a similar result for ADQC and we present 
a transformation between ADQC and MBQC that preserves depth.

\subsection {GBQC and ADQC}

The question of translating GBQC circuits into MBQC patterns and
vice versa has been addressed before in~\cite{BK06} and it can be
directly adapted for ADQC as well. In fact the
universality proof of ADQC already presents a method of translation of
a circuit into ADQC: (I) Rewrite the given circuit in terms of the universal gates set of
  $J(\al)$ and $\widetilde{\ctR Z}$; (II) Replace each gate with its corresponding ADQC pattern (equations
  \ref{e-Jpat3} and \ref{e-ctrZpat}); (III) Perform the standardisation procedure.

\begin{figure}[h]
\begin{center}
\includegraphics[scale=0.4]{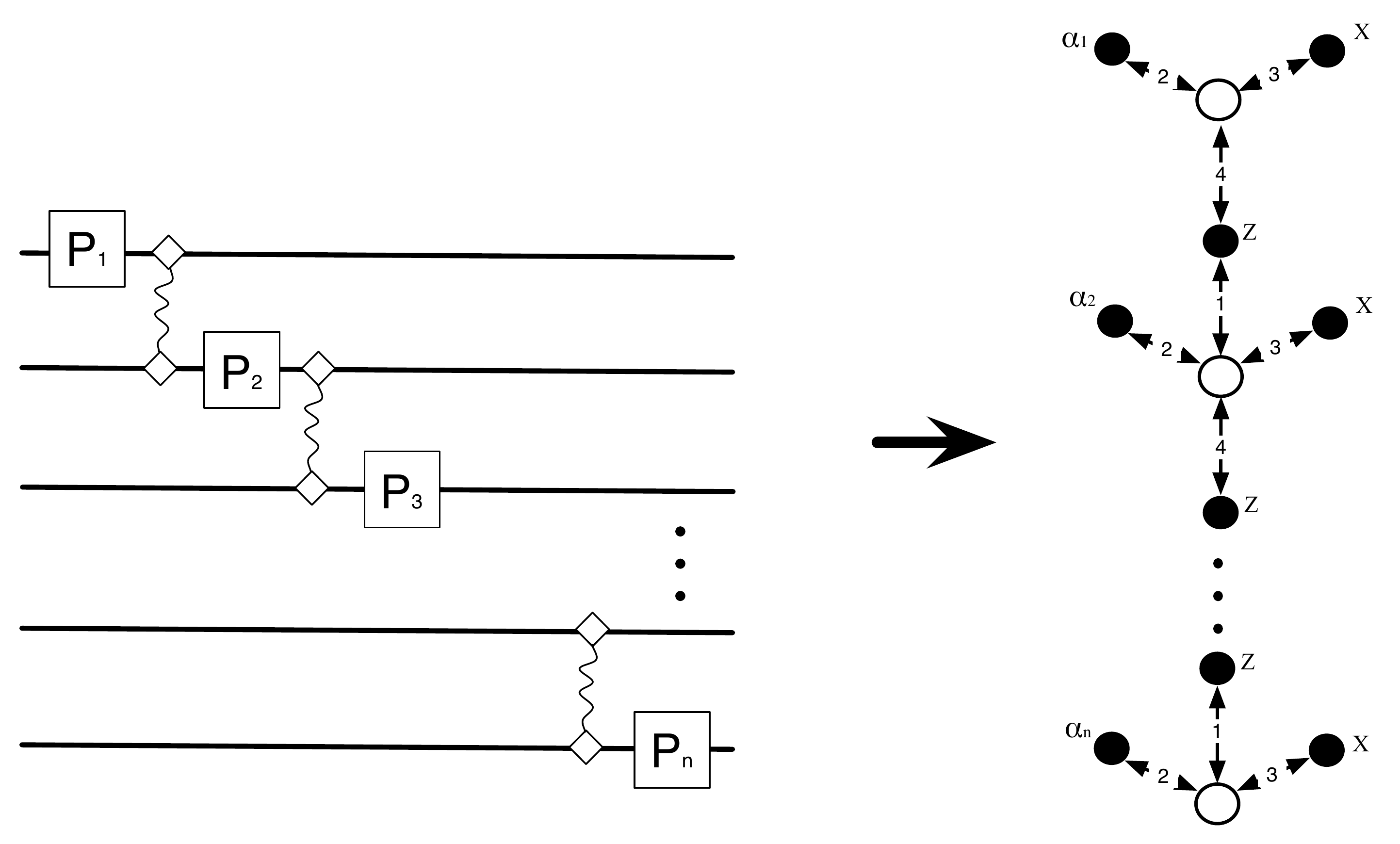}
\caption{A ladder structure circuit with the binary $\widetilde{\ctR Z}$ and the unitary $P(\al_i)$ gates, together with the
  corresponding open twisted graph state obtained via gate by gate translation. The edge labels present the edge colouring and the vertex labels are the measurement angles, where $X$ and $Z$ stands for Pauli measurement.}
\label{f-lad}
\end{center}
\end{figure}

The above construction cannot be used in reverse as the edge colouring
order might lead to a circuit with an acausal loop. However it is
possible to keep all the auxiliary qubits to avoid creating loops in
the resulting circuit. The scheme is simply based on the well-known
method of coherently implementing a measurement \cite{BK06}. It is also easy to prove, in a similar way as in \cite{BK06}, that the
translation from a GBQC circuit into an ADQC pattern will never
increase the depth as the edge colouring number of the obtained twisted graph state will be upper-bounded by the depth of the original
circuit. More importantly, we present an example where the depth
decreases exponentially. Consider the ladder structure of the circuit
in Figure \ref{f-lad} which has depth $n$. This circuit, through the
introduced construction, will be translated into a pattern with the
twisted graph state shown in Figure \ref{f-lad} which has constant
depth $4$. This is due to the fact that the preparation depth for any ADQC pattern by definition, is upper bounded by the edge colouring of the graph, which in this case is equal to 4. On the other hand the Pauli $X$ measurements on the right most qubits, break the dependency chain between measurement \ie~an $\al_i$ measurement depends only on the result of some Pauli measurements. Hence one can find a flow for this twisted graph state where all the Pauli measurements are in the first layer of the corresponding partial order and all the non-Pauli measurements are in the final layer and computation depth which is equal to the flow depth is 2. It is easy to verify that the same logarithmic depth separation result between GBQC and MBQC obtained for the parity function~\cite{BK06} is also valid for the
case of GBQC and ADQC.

\subsection{MBQC and ADQC}
\label{sec:trans-mbqc}

As mentioned before there exists a compositional embedding from MBQC
patterns with flow into GBQC and vice versa, and together with the
construction of the last subsection one can obtain an embedding
between ADQC and MBQC for patterns with flow. However in view of
parallelism, it is interesting to find such an embedding directly by
presenting the correspondence between twisted open graph states and
open graph states.

The following equation relates the two resources but it is only valid
for degree-one ancilla qubits
\EQ{\label{e-swap1} 
\etil a s N_a^{\ket +}=\text{SWAP}_{as}\, \et a s N_a^{\ket +}.
} 
This equation and the
next one are in fact the reason behind the chosen name for this class
of states as one can recover a graph state from them by applying the
appropriate sequence of twist (SWAP) operators. In order to handle the
degree-two ancilla qubits we will use the following pattern equations
\EQ{ \nonumber \mfr {\widetilde{\ctR Z}} &=& \cx s {m_a} M^{Z}_a
  \etil a{s'} \etil as \\ \nonumber &=& \cx s {m_a} M^{Z}_a H_a H_{s'}
  \et a{s'} \etil as \\ \nonumber &=& \cx s {m_a} M^{X}_a H_{s'} \et
  a{s'} \etil as \\ \label{e-swap2} &=& \cx {s'} {m_b} \cx s {m_a}
  M^{X}_a M^{X}_b \etil b{s'} \et a{s'} \etil as } 
In the new pattern for $\nonumber \mfr {\widetilde{\ctR Z}}$ both instances of $\etil a s$ can be replaced using Equation \ref{e-swap1}. 

\EQ{ \nonumber 
\mfr {\widetilde{\ctR Z}} &=&  \cx {s'} {m_b} \cx s {m_a} \; M^{X}_a M^{X}_b \; \text{SWAP}_{bs'} \et b {s'} \; \et a{s'} \; \text{SWAP}_{as} \et a s \\
\label{e-swap3} &=& \cx {b} {m_{s'}} \cx a {m_s} \; M^{X}_s M^{X}_{s'} \; \et b {s'} \et s{s'}  \et a s
}

Therefore we can replace any pattern over a given twisted graph state where degree-two
vertices are measured with Pauli $Z$, into a pattern over a graph
state obtained through the above manipulations of the $\etil{a}{s}$
edges.

The other direction of translation, \ie~from graph sates to twisted graph states can be obtained from the following general equation.
\EQ{ \nonumber 
\et s {s'} &=&  H_{s'} \;\; H_s \;\; \etil s {s'} \\\nonumber
&=& H_{s'} \;\; H_s \;\; \cx s {m_a} M^{Z}_a \etil a{s'} \etil as \\\nonumber
&=& \cx {s'} {m_{b'}} M^{X}_{b'} \etil {s'} {b'} \;\; \cx s {m_b} M^{X}_b \etil s b \;\; \cx s {m_a} M^{Z}_a \etil a{s'} \etil as \\\label{e-general}
&=& \cx {s'} {m_{b'}} \cx s {m_b+m_a} \;\; M^{X}_{b'} M^{X}_b  M^{Z}_a \;\; \etil {s'} {b'} \etil a{s'} \etil s b \etil as
}
Not that the resulting twisted graph state might not be unique as it depends on the order of re-write rules application. However for the special class of graph state with flow \cite{Flow06}, which are still universal for MBQC, we present a unique translation that is also more efficient, \ie~uses fewer ancilla qubits. Consider a graph state $(G,I,O)$ with flow $(f, \preceq)$, see Appendix \ref{ss-mbqc} for definitions. Denote the orbit of $f$ on $i \in I$ with $F_i=\{v\in V | \exists n : f^n(i)=v\}$ and let $o_i$ be the unique element in $O\cap F_i$. Due to the property of the flow \cite{Flow06}, every $v\in V$ belongs to a unique $F_i$ so we can define the rank of $v$ to be the unique value $r(v)$ such that $f^{r(v)}(i)=v$\,. The corresponding twisted graph state $G_t(S,A)$ is constructed by the following steps, where the system qubits in $G_t$ are the input qubits in $G$.
\begin{itemize}
\item[1.] For any $i\in S$, add $r(o_i)$ many degree-one ancilla qubits to $A$ and attach them to $i$, and set the edge labels $1, \cdots, r(o_i)$.
\item[2.] For any non-flow edges in $G$ between $a\in F_i$ and $b\in F_j$ add a degree-two ancilla qubit $q$ to $A$ with edges $\etil i q$ and $\etil j q$. Set the edge label of $\etil i q$ and $\etil j q$ to be $r(a)+1$ and $r(b)+1$ (if $a$ and $b$ have the same rank make one of the edge label bigger). Finally adjust the labelling of the added edges in the first step in accordance of these new labels.
\end{itemize}

It is a straightforward but cumbersome computation to show that the above translation from MBQC into ADQC is depth preserving so we omit the details of the proof. We leave as an open question whether a general depth preserving translation between MBQC and ADQC can be constructed. 

\section{Discussion}\label{s-dis}

ADQC presents significant advantages over GBQC for particular physical
implementations. By isolating the system memory from measurement and
state preparation, the physical layout of a quantum computer can be
optimised. Potentially decoherent read-out mechanisms can be located
away from the memory. Since only a fixed two-qubit unitary gate has to
be implemented between memory and ancilla qubits, this simplifies
considerably the construction, characterisation, control and operation
of the computer. Control line clutter can be reduced, due to relaxing
the need to implement single qubit rotations and measurement on the
memory, simplifying the architecture of the computer as well as
minimising the possibility of cross-talk. The choice of physical
qubits for memory and ancilla can also be optimised. Memory qubits can
be chosen for long coherence times at the expense of being static,
whilst ancilla qubit can be chosen for mobility, and ease of
initialisation and measurement.

Natural candidate systems consist of an array of static qubits
addressed by flying qubits or a ``read/write'' head. For example, an
optical lattice of neutral atoms can be addressed by a separately
controlled atom~\cite{YC2000,CDJWZ2004}. Since it is difficult to
individually address with lasers a single site of a fully filled
optical lattice, the read-write head would interact with selected
sites and can be used as the ancilla. A similar idea can be applied to
neutral atoms trapped in dipole trap arrays~\cite{DVMMME2002}
controlled by an atom in an optical
tweezer~\cite{BTMGMSLJMBG2007}. Similarly in~\cite{CZ2000}, ions in an
array of micro traps would be manipulated by a single ion read-write
head. Alternatively in~\cite{GVTE2000}, the role of the memory is
played by the quantised electromagnetic field in an array of cavities
whilst Rydberg atoms traversing the cavities act as ancillas (similar
ideas are contained in~\cite{BV2006}). All of the above schemes
possess a natural $\ctR Z$ (dispersive) interaction between the
ancilla and system qubits. For universality, either an additional
Hadamard operation on the system should be incorporated into this
interaction (e.g. through the pushing laser in~\cite{CZ2000}), or an
effective SWAP operation be found in conjunction with the
Controlled-Z. The latter could be achieved through cold collisions
between ancilla and system~\cite{CEMW2002,EMYSBBL2002}.
 
A system which may prove particularly amenable to our model is discussed in~\cite{GCHH2004}. Here, the nuclear spin of a single dopant
atom in isotopically pure silicon plays the role of a memory qubit
which can be controllably coupled via the hyperfine interaction to an
electron spin which acts as an ancilla qubit. Nuclear spins can be
very well isolated from the environment, as well as the state
preparation and measurement areas. Electrons can be rapidly
transported around the computer using charge transport via adiabatic
passage (CTAP), this avoids the issue of swapping nuclear spin
states, as in the original Kane proposal~\cite{Kane1998}, which can
lead to a reduction in fault tolerance. An issue here is making sure
that the interaction between electron and nuclear spins is of the
correct form as to allow conditional unitary dynamics. An interaction of the form $\ctR Z$ followed by SWAP gate can be achieved by using the
method presented in~\cite{LdV1998} with only the Heisenberg
interaction between ancilla and system and local operations on the
ancilla itself.
 
A natural interaction which is also suitable for ADQC is the
XY-Hamiltonian which be easily turned into the $\ctRZSWAP$ 
gate~\cite{SS2003} (equivalent to the ISWAP in the aforementioned
reference). In~\cite{MPL2001}, a natural XY-interaction between
nuclear spins, as in the Kane proposal, is mediated by a 2D electron
gas in the Quantum Hall regime. The XY-interaction also naturally
occurs between quantum dots coupled by a cavity~\cite{IABDLSS1999} or
superconducting qubits~\cite{SFPE2000,LOMM2001}. Here, the memory and
ancilla qubits are of the same species.

\end{document}